\documentclass[final,3p,times,twocolumn]{elsarticle}

\usepackage{amssymb}
\usepackage{graphicx}
\usepackage[abs]{overpic}
\usepackage{booktabs} % useful for tabs
\usepackage{amsmath}
\usepackage{url}
\usepackage{lineno}
\usepackage{color}
\input babarsym	% official BABAR symbol definitions
\def\svt {SVT}
\def\dch {DCH}
\def\dirc {DIRC}
\def\emc {EMC}

\journal{Nuclear Instruments and Methods in Physics Research A}

\begin{document}

%\begin{flushleft}
%\mbox{\normalsize {SLAC-PUB-15095} }
%\end{flushleft}

\nopagebreak[0]

\begin{frontmatter}

\title{\hspace{-7cm}\normalsize{SLAC-PUB-15095}\vspace{1cm}
    \Large{ \newline Track Finding Efficiency in \babar}}

\author[ohio]{T.~Allmendinger}
\author{B.~Bhuyan\corref{}\fnref{bb}}
\ead{bhuyan@iitg.ernet.in}
%\author[bb]{B.~Bhuyan}
%\corref{Bipul Bhuyan}
\author[lbl]{D.~N.~Brown}
\author[uvic]{H.~Choi}
\author[rostock]{S.~Christ}
\author[infn]{R.~Covarelli}
\author[orsay]{M.~Davier}
\author[mainz]{A.~G.~Denig}
\author[mainz]{M.~Fritsch}
\author[mainz]{A.~Hafner}
\author[uvic]{R.~Kowalewski}
\author[ucr]{O.~Long}
\author[orsay]{A.~M.~Lutz}
\author[mm]{M.~Martinelli}
\author[slac]{D.~R.~Muller}
\author[uvic]{I.~M.~Nugent}
\author[princeton]{D.~Lopes~Pegna}
\author[sc]{M.~V.~Purohit}
\author[mainz]{E.~Prencipe}
\author[uvic]{J.~M.~Roney} 
\author[pd]{G.~Simi}
\author[novo]{E.~P.~Solodov}
\author[princeton]{A.~V.~Telnov}
\author[princeton]{E.~Varnes}
\author[rostock]{R.~Waldi}
\author[notre]{W.~F.~Wang}
\author[sc]{R.~M.~White}

%\author{}
\address[ohio]{Ohio State University, Columbus, Ohio 43210, USA}
\address[bb]{Indian Institute of Technology Guwahati, Assam, 781 039, India}
\address[lbl]{Lawrence Berkeley National Laboratory and University of California, Berkeley, California 94720, USA}
\address[uvic]{University of Victoria, Victoria, BC, V8W 3P6, Canada}
\address[rostock]{Universit\"at Rostock, D-18051 Rostock, Germany }
\address[infn]{INFN Sezione di Perugia, Dipartimento di Fisica, Universit\`a di Perugia, I-06100 Perugia, Italy }
\address[orsay]{Laboratoire de l'Acc\'el\'erateur Lin\'eaire, IN2P3/CNRS et Universit\'e Paris-Sud 11, Centre Scientifique d'Orsay, B.~P. 34, F-91898 Orsay Cedex, France }
\address[mainz]{Johannes Gutenberg-Universit\"at Mainz, Institut f\"ur Kernphysik, D-55099 Mainz, Germany }
\address[ucr]{University of California at Riverside, Riverside, California 92521, USA}
\address[mm]{INFN Sezione di Bari, I-70126 Bari, Italy}
\address[slac]{SLAC National Accelerator Laboratory, Stanford, California 94309 USA }
\address[princeton]{Princeton University, Princeton, New Jersey 08544, USA }
\address[sc]{University of South Carolina, Columbia, South Carolina, 29208, USA}
\address[pd]{Universit\'a di Padova, I-35131 Padova, Italy}
\address[novo]{Budker Institute of Nuclear Physics, Novosibirsk 630090, Russia }
\address[notre]{University of Notre Dame, Notre Dame, Indiana 46556, USA }

%\address{}

\begin{abstract}
We describe several studies to measure the charged track
reconstruction efficiency and asymmetry of the \babar\ detector.  The
first two studies measure the tracking efficiency of a charged particle 
using $\tau$  and initial state radiation decays. The third uses the $\tau$
decays to study the asymmetry in tracking, the fourth measures the 
tracking efficiency for low momentum tracks, and the last measures the 
reconstruction efficiency of $K^0_S$ particles. The first section also 
examines the stability of the measurements vs \babar\ running periods. 
\end{abstract}

\begin{keyword}
\babar \sep tracking \sep efficiency


\end{keyword}

\end{frontmatter}

%\clearpage
%\pagebreak
\linenumbers

\section{\bf Introduction}
The \babar\ experiment operated from 1999 to 2008 at the
\pep2\ asymmetric \epem\ collider at the SLAC National Accelerator
Laboratory.  \babar\ was designed to study CP violation and other rare
decays in flavor physics from events produced at or near the $\Upsilon$
resonances, from 9.46~GeV to over 11~GeV.  A critical requirement for
meeting {\babar}'s science goals was the ability to efficiently and
accurately detect stable charged particles, or {\em tracks}, produced in
\epem\ collisions.  Many analyses performed at \babar\ require a precise
estimate of the track finding efficiency, as input for measuring the
absolute or relative rate of the physics process being studied.

In this paper, we present the algorithms and methods used in \babar\ to
estimate the track finding efficiency.  To cover the range of particle
momenta and production environments relevant to most \babar\ analyses, a
number of methods are used.
%By comparing the effiiency estimates from the different methods
%in their regions of overlap, we obtain a cross-check on the results, and a
%verification of the error we assign to the results. 
To compute the tracking efficiency from data alone, these methods rely
on special data samples, where additional constraints can be applied.
The primary efficiency result is computed using $\epem \rightarrow
\taup\taum$ events, which can be cleanly isolated in the \babar\ data
sample, and which have a simple topology.  To cross-check this result,
we independently measure the tracking efficiency using radiative $\epem
\rightarrow \pipi\pipi\gamma_{ISR}$ events, where $\gamma_{ISR}$ is an 
initial state radiation (ISR) photon, which can be constrained
kinematically.  To study the reconstruction efficiency of low
momentum tracks, we use $D^{*\pm} \rightarrow
D^{0} \pi^{\pm}$ decays.
%%, where spin effects allow a determination of the ``slow'' pion efficiency.  
We also present a dedicated study of the
efficiency to reconstruct $K_S^0 \rightarrow \pipi$, whose daughter
tracks can have a different efficiency due to their displacement
from the primary event origin.

The strategy for the $\tau$-based
and $\epem \rightarrow \pipi\pipi\gamma_{ISR}$ track reconstruction efficiency
measurements is to use
charge conservation and kinematics to deduce the
existence of a track, given a subset of detected tracks in well-defined events.
The efficiency analyses based on
$D^0$ decays and for the $K^0_S$ efficiency study use a statistical
approach, using properties of momentum distributions which will be
described below.  Systematic errors are estimated using internal self-consistency
measures and by comparing different efficiency analysis techniques.

The \babar\ detector geometry, material, and sensor response functions
have been accurately modeled in a detailed simulation based on the
Geant4 \cite{g4} framework.  The output of the \babar\ simulation is
processed using the same reconstruction algorithms as applied to data,
and the results have been found to be very similar to what we see in
data.  By using accurate computer models of the physics processes
relevant at \babar\ energies \cite{generators,Lange:2001uf}, we are able to generate
equivalent samples of simulated data as used in nearly all \babar\
analyses, including the tracking efficiency analysis.
%select for measuring the tracking efficiency,
%and thereby repeat the efficiency analysis on simulated data nearly exactly as we perform
%it on data.  
\babar\ has therefore adopted the strategy of estimating the tracking
efficiency relative to that observed in the simulation, which 
simplifies the application of the tracking efficiency results in
analysis.  As will be shown in the following sections, for most of the studies, the tracking
efficiency found in data agrees within errors with the efficiency found
in simulated data.  This allows the result of the tracking efficiency
measurement to be used in analysis simply by propagating the appropriate
systematic errors on the tracks involved to the simulation estimate of
the analysis signal reconstruction efficiency.  This strategy has been
used in many scientific \babar\ publications. However, for analyses involving a 
$K_S^0$, a correction is required in the MC for its daughter 
reconstruction efficiency which depends on the kinematics of the decay of interest.

\section {\bf  \babar\ Detector and Data Sample}
\label{DetectorandDS}
The \babar\ detector is a multi-purpose device designed to
simultaneously measure many properties of the multiple particles
produced in \epem collisions near the $\Upsilon$ resonances, as
described in detail in \cite{babar_nim}.  Charged particles are
identified in a Silicon Vertex Tracker (\svt), and a Drift Chamber (\dch),
which are surrounded by a superconducting solenoid that generates an
approximately uniform 1.5 Tesla magnetic field inside the
sensitive volumes of these detectors.

The \svt ~consists of five layers of double-sided silicon strip
detectors covering the full azimuthal range and the laboratory frame
polar angle ($\theta_{lab}$) range $20.1^\circ < \theta_{lab} < 150.2^\circ$
~\cite{babar_nim}. The intrinsic resolution of individual
\svt\ position measurements of particles which traverse it varies between
$10\mum$ and $30\mum$, depending on the incident particle direction and
the readout view.  The \dch ~consists of 7,104 hexagonal drift cells, which are approximately
1.9 cm-wide and 1.2 cm-high, made up of one sense wire surrounded by six field wires. The sense wires are
$20\mum$ gold-plated tungsten-rhenium, the field wires are $120\mum$ and $80\mum$ gold-plated aluminium.
The cells are arranged in 40 cylindrical layers and the layers are grouped by four into ten {\em super-layers}
extending from roughly 25 cm to 80 cm in the transverse direction, with full coverage over the range
$24.8^\circ< \theta_{lab} < 141.4^\circ$, and partial coverage over the range
$17.2^\circ< \theta_{lab} < 152.6^\circ$.  
%40 layers of alternating 4-layer {\em super-layers} of axial and stereo wires 
The intrinsic resolution of individual measurements of track position in the \dch\
varies between $100\mum$ and $200\mum$, depending on the track position
and angle relative to the wire, with an average resolution of $150\mum$.

The \babar\ detector includes a dedicated charged particle
identification (PID) device based on detection of internally reflected
Cherenkov radiation (\dirc), and a Cesium-iodide crystal electromagnetic
calorimeter (\emc) for identifying electrons and photons.  The steel for
the solenoid magnet flux return is instrumented with position-sensitive
chambers, which produce distinctive signatures from passing muons and
pions.  \babar\ estimates the species of charged particles using a
combination of information from these devices, plus the specific
ionization (\dedx) measured in both the \svt\ and \dch.  By studying
the response of these systems to high-purity control samples,
% Probability Density Functions (PDFs) are measured, and combined to form
likelihood functions describing a track's consistency with each
of the 5 charged particle species ($e^\pm, \mu^\pm, \pi^\pm, K^\pm$, and $p^\pm$) directly observable
in the \babar\ tracking system are defined.  Samples of specific particle
species of varying efficiency and purity are selected by cutting on appropriate
likelihood ratios.

The results presented in this paper are based on the full \babar\ data sample, 
collected in seven distinct periods,  Runs 1-7. 
Runs 1-6 correspond to data collected with a
center-of-mass (CM) collision energy near or at the $\Upsilon\left(4S\right)$
resonance and Run 7 corresponds to the data collected with a CM
collision energy at the $\Upsilon\left(3S\right)$ and 
$\Upsilon\left(2S\right)$ resonances. 

\section {\bf \babar\ Track Reconstruction Algorithms}
Tracks are reconstructed in \babar\ using a combination of several
algorithms.  Tracks with transverse momentum above roughly 150 MeV/c are
principally found in the \dch.  Track {\em segments} are identified as
contiguous sets of hits in a super-layer having a pattern consistent with
coming from a roughly radial track.  Segments are linked using their
position and angle to form a track candidate.  Track candidates are fit
to a helix, which is used to resolve the left-right ambiguity, and to
remove {\it outlier} hits.  The candidate is kept if at least 20 \dch
~hits remain.  Tracks with large impact parameter are found in the \dch\
using a less restrictive algorithm.  Tracks found in the \dch\ are fit using a
Kalman filter \cite{kalman_chep} fit, which accounts for material
effects and corrects for magnetic field inhomogeneities.  The Kalman
filter track fit is extrapolated inwards, and \svt\ hits consistent with
the extrapolated track position and covariance are added.
%position within $5 \sigma$ of the extrapolated fit covariance are
%selected.  Patterns of \svt\ hits consistent with the passage of a high-momentum track,
%including possible missing hits, are identified, and the $\chi^2$ of the resulting Kalman
%filter fit of both \dch\ and \svt\ hits is
%computed.  A fixed {\em penalty term} for each missing hit not corresponding to a
%known dead regions of the \svt\ is added to the fit $\chi^2$, and the pattern of \svt\ hits with the lowest
%total $\chi^2$ is selected.

Tracks with low transverse momentum are found principally in the \svt\,
using hits not already associated with tracks found in the \dch. Sets of
four or more {\em $\phi$} hits (which measure the position in the plane
transverse to the beam direction), in different layers of the \svt\ and 
consistent with lying on a circle, are selected.  Hits in the orthogonal
($z$) view of the same wafers as the $\phi$ hits are then added to form
three-dimensional track candidates.  Candidates with at least 8 hits are
selected, and fit using the Kalman filter.  Additional tracks are found
in the \svt\ using space points constructed from pairs of $\phi$ and $z$
hits not already used in other tracks.  Sets of at least 4 space points
consistent with a helix fit are selected as tracks.  \dch\ hits are
added to tracks found in the \svt\ in a procedure analogous to how
\svt\ hits are added to tracks found in the \dch.

After all the tracks in an event are found, they are filtered to remove
duplicate tracks due to hard scattering in the material separating the \svt\
and the \dch, decays in flight, or pattern recognition errors in the
\dch, where stereo and axial hits generated by a single particle are
sometimes reconstructed as separate tracks.  A final pass to remove
inconsistent hits and to add individual hits missed in the pattern
recognition is then performed using the Kalman filter fit.

The resultant set of tracks is referred
to as {\em Charged Tracks} (CT).  A  {\em Good Tracks} (GT) 
subset of tracks, with a higher probability of originating from the primary
\epem interaction, is selected from these.  The GT selection
requires the impact parameter with respect to the average interaction point 
be less than 1.5 cm in the transverse direction, and less than 2.5 cm along the
magentic field (z) direction. Analyses at \babar~generally
use either the CT or the GT track selection, and the
tracking efficiency studies described in this note are performed independently
for both.

\section{\bf Tau31 Tracking Efficiency Study}
\label{sec:tau31eff}

The efficiency of charged track reconstruction at \babar~is determined using
$e^{+}e^{-}\to \tau^{+}\tau^{-}$ events.  With over 430
million $\tau$ pair events collected at \babar, $\tau$ decays provide an
opportunity to make a precision measurement of the tracking efficiency.
At the CM energies produced at \babar, $\tau$ decays
are an ideal candidate for measuring the tracking efficiency because
they have a momentum and angular distributions of tracks that are
similar to those from decays of D and B mesons.  Decays of $\tau$
leptons have a high track density due to the initial boost,
~$\beta \sim0.94c$, of the $\tau$ leptons, while the total track
multiplicity is low.  The $\tau$ lepton has a life-time of $(290.6\pm
1.0)\times 10^{-15}s$, which results in a transverse flight length of
$200~\mu m$ at the \babar\ CM energies, a value
that is slightly larger than the beam spot size but small enough not to
impact the tracking efficiency.

The tracking
efficiency is measured using $e^{+}e^{-}\to \tau^{+}\tau^{-}$ events
in which one $\tau$ lepton decays leptonically via
$\tau^{\pm}\to\mu^{\pm}\nu_{\mu}\nu_{\tau}$, and the other
$\tau$ lepton decays semi-leptonically to 3 charged hadrons via $\tau^{\mp}\to
h^{\mp}h^{\mp}h^{\pm}\nu_{\tau}\ +\ \ge 0$ neutrals (excluding $K^0$),
referred to as {\em Tau31} events.
The tracking efficiency is measured using the
3-prong $\tau$ decays.  The branching ratio of
$\tau^{\pm}\to\mu^{\pm}\nu_{\mu}\nu_{\tau}$ and 3-prong $\tau$ decays
are $(17.36\pm 0.05)\%$ and $(14.56\pm 0.08)\%$~\cite{pdg} respectively,
so that Tau31 events constitute over 5\% of the total.
The $\tau$ pair candidates are selected by requiring an isolated muon track,
plus at least two other reconstructed tracks consistent with being hadrons.
Events are selected in two overlapping channels; those where two of the hadrons
have the same charge (``same-sign''), and those where two of the hadrons have opposite
charge (``opposite-sign'').
Requiring a muon track is an essential part of suppressing
non-$\tau$ backgrounds: radiative Bhabha events where the photon
interacts with the detector material producing an $e^{+}e^{-}$ pair
(conversion), $\gamma$-$\gamma$ events, and $q \bar{q}$ events.  Charge
conservation infers the existence of the fourth track. 

\begin{figure}
\centering
\includegraphics[width=0.35\textwidth]{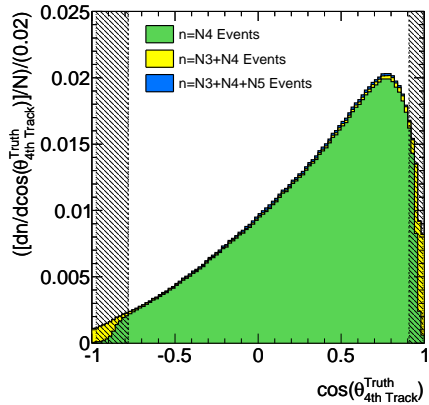}
\caption[The true $cos(\theta)$ of the fourth track in the Tau31 study]{
  The true $cos(\theta)$ in the laboratory frame for the fourth track
  for the selected opposite-sign and same-sign MC
  events. $N=N_3+N_4+N_5$ is the number of selected same-sign and
  opposite-sign events. $N_4$ is the number of events where the fourth
  track is found for the CT definition, $N_3$ is the number of events
  where the fourth track is not found for the CT definition and $N_5$ is
  the number of events where two CT candidates are found for the fourth
  track. The dotted lines indicate the outer edge of the tracking
  detectors, while the dashed lines indicate the edge where the full
  detector coverage begins. The region in between where there is partial
  coverage is indicated by the shading.
\label{fig:tau31:Accpttheta}}
\end{figure}

\begin{figure}
\centering
\includegraphics[width=0.35\textwidth]{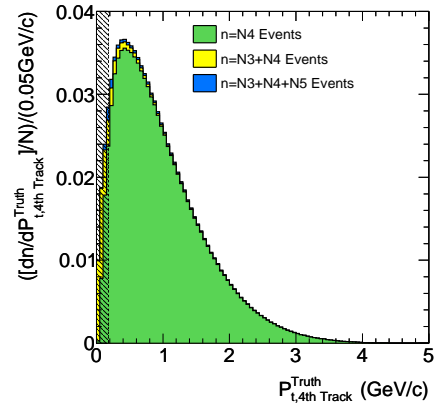}
\caption[The true $p_{t}$ in the laboratory frame of the fourth track in the Tau31
  study]{ The true  $p_{t}$ for the fourth track for 
the selected opposite-sign and same-sign MC events. 
  $N=N_3+N_4+N_5$ is the number of selected
 same-sign and opposite-sign
  events, $N_4$ is the number of events where the fourth track is
  found for the CT definition,
$N_3$ is the number of events where the fourth track is not found for
  the CT definition and
$N_5$ is the number of events where two CT candidates are found for
  the fourth track.  The tracks in the shaded region do not reach the
  outer edge of the DCH.
\label{fig:tau31:Accptpt}}
\end{figure}

The tracking efficiency $\epsilon$ is defined by

\begin{equation}
\epsilon \times A = \frac{N_4}{N_3 + N_4}
\label{Tau31eff}
\end{equation}

\noindent 
where $A$ is the geometric
acceptance of the
fourth track
constrained by the $\tau$ pair kinematics and the selection criteria
of the Tau31 sample, $N_4$
is the number of events where the fourth track is found, and $N_3$ is
the number of events where the fourth track is not found.
The geometric acceptance of the \babar\ detector
for a uniform $cos(\theta)$ distribution is $\sim83$\%.
In figures
\ref{fig:tau31:Accpttheta} and \ref{fig:tau31:Accptpt}, the geometric
acceptance of the detector is plotted for simulated events as a function of the 
polar angle ($\theta$) and the transverse momentum
($p_{t}$) of the fourth track, respectively. These
figures demonstrate the
limited angular acceptance of the detector, and the poor acceptance for low momentum tracks.

\subsection{Monte Carlo Samples}

$\tau^{+}\tau^{-}$ pair events are simulated with higher-order radiative
corrections using the {\tt KK2f} Monte Carlo (MC) generator~\cite{kk}
with $\tau$ decays  simulated with
{\tt Tauola}~\cite{tauola,photos}. The simulated Standard Model backgrounds
include: $b\bar{b}$; $c\bar{c}$; $s\bar{s}$; $u\bar{u}$; and
$\mu^{+}\mu^{-}$
events~\cite{Lange:2001uf,kk,tauola,photos,jetset}.
The number of simulated background events is comparable to the number
expected in the data,
with the exception of Bhabha and two-photon events, which are not
simulated. Bhabha
and two-photon events backgrounds are studied with control samples.
The detector simulation and reconstruction of the MC events is
described in Section \ref{DetectorandDS}.  

\subsection{Event Selection}

We require the events to have a minimum of three GT and a
maximum of five CT tracks.  Events with \KS are removed, where the \KS
candidate is defined as having two oppositely charged tracks with an
invariant mass within 10~\mev of the $\KS$ mass~\cite{pdg}, a vertex
displaced more than 2~mm from the beam-spot and a vertex fit $\chi^{2}$ probability
of more than 1\%.  The three GT tracks
%originating from the beam
are required to have $p_{t}>100$\mev.  To remove any remaining duplicate
tracks, the three GT tracks are required to satisfy an isolation cut in
$\theta$, $\phi$ and momentum by 0.1 rad, 0.1 rad and 0.4 \gev,
respectively.  One of the three GT tracks must be more than 120 degrees
from the other track. %Figure \ref{fig:trkeff4:betamupi} shows the cosine
%of the angle between the muon and the closest identified pion
%($cos(\theta_{\mu\pi})$). 
This isolated track must satisfy a tight muon PID selection.
%The muon tag is essential to suppress background from radiative Bhabha
%events where the photon converts producing a $e^{+}e^{-}$ pair.
At least two of the other
%well reconstructed
tracks are required to be identified
as pions, by being inconsistent with a loose electron PID
selection.
%\begin{figure}
%\centering
%\includegraphics[width=0.45\textwidth]{costhetalpipi.eps}
%\caption[The isolation angle between the muon and the pions  ]{
%The cosine of the angle between the muon and the closest identified
%pion ($cos(\theta_{\mu\pi})$) with all other selection criteria
%applied for about 15\% of the \babar\ Dataset.
%The points represents
%the data, the empty histogram represents the
%3 prong $\tau$ decays, the light shaded
%histogram represents the other $\tau^{+}\tau^{-}$ MC, the medium dark
%histogram represent the $\mu^{+}\mu^{-}$ MC and the dark histogram
%represents the $q\bar{q}$ MC.  The background contamination in these
%samples is small.
%  \label{fig:trkeff4:betamupi}}
%\end{figure}

\begin{figure}
\centering
\includegraphics[width=0.45\textwidth]{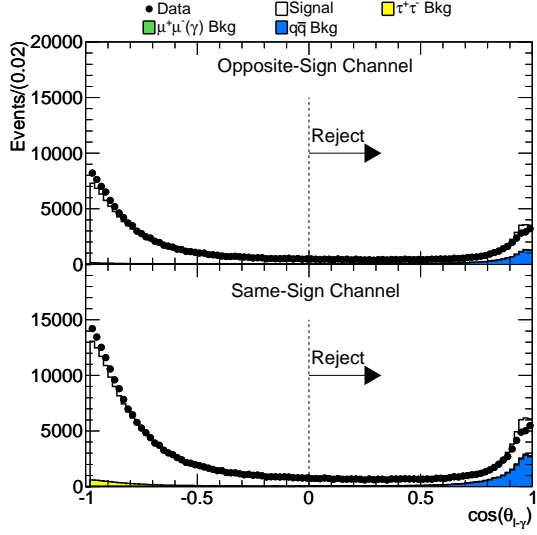}
\caption[The isolation angle between the muon and the pions  ]{
The cosine of the angle between the muon and the closest identified
photon ($cos(\theta_{\mu\gamma})$) with all other selection criteria
applied for about 15\% of the \babar\ data sample. The points represent
the data, the empty histogram represents the
3 prong $\tau$ decays, the light shaded
histogram represents the other $\tau^{+}\tau^{-}$ MC, the medium dark
histogram represent the $\mu^{+}\mu^{-}$ MC and the dark histogram
represents the $q\bar{q}$ MC.  The background contamination in these
samples is small.
  \label{fig:trkeff4:betamugamma}}
\end{figure}

%The events are classified as either ``same-sign'' or
%``opposite-sign''. 
%The ``same-sign'' is the loose selection while the
%``opposite-sign'' uses the selection of the $\rho$ to produce a tight selection.
For the ``same-sign'' channel ($\tau^{\pm}\rightarrow\pi^{\pm}\pi^{\pm}
X^{\mp}\nu_{\tau}$, where $X^{\mp}$ is the unidentified $4^{th}$
track), we require 
%0.3\gev$<M_{\pi^{\pm}\pi^{\pm}}<M_{\tau} + $0.05\gev 
0.3\gev$<M_{\pi^{\pm}\pi^{\pm}}<M_{\tau}$ to ensure that the charged pions 
are consistent with coming from a $\tau$ lepton decay.  
For the ``opposite-sign'' channel 
($\tau^{\pm}\rightarrow \pi^{\pm}\pi^{\mp}X^{\pm}\nu_{\tau}$), we require 
$|M_{\pi^{\pm}\pi^{\mp}}-M_{\rho}|<100$\mev to ensure that the charged pions 
are consistent with coming from a $\rho$ meson.
This produces a loose selection for the ``same-sign'' channel and a
tight selection for the ``opposite-sign'' channel.
An event can be selected in either or both channels.
In the case where more than one
same-sign or opposite-sign pion pairing is possible, the
pair with the highest laboratory frame $p_t$ is selected.

To remove $q\bar{q}$ backgrounds, events with neutral particles with an
energy greater than 0.5 \gev that are within 90 degrees from the muon track are
removed. Figure \ref{fig:trkeff4:betamugamma} shows the cosine of the
angle between the muon and the photon ($cos(\theta_{\mu\gamma})$). To suppress
radiative di-muon and Bhabha backgrounds with conversions, the muon track must
have a CM momentum, ($p_{\mu}^{CM}$) less than 80\% and
greater than 20\% of $\sqrt{s}/2$, where $\sqrt{s}$ is the beam CM energy. To further
reduce the non-$\tau$ backgrounds, the polar angle of the system of
charged particles,
the $\mu$-$\pi\pi$ system, in the CM frame must satisfy
$|\cos(\theta_{\mu-\pi\pi})|<0.8$, with the net transverse momentum of the
$\mu$-$\pi\pi$ system being more than 0.3 \gev.

\begin{figure}
\centering
\includegraphics[width=0.45\textwidth]{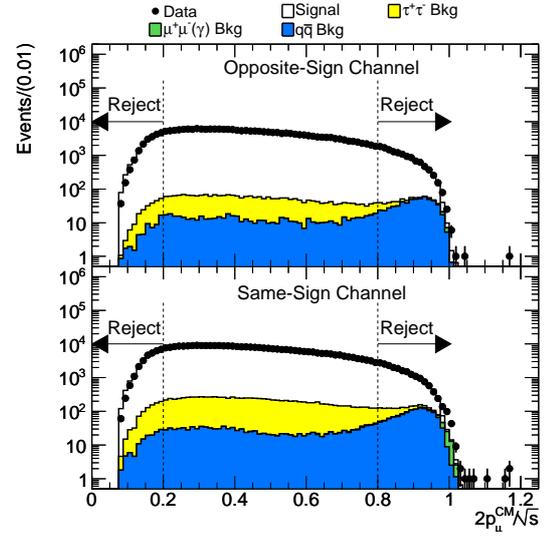}
\caption[The $2p_{\mu}^{CM}/\sqrt{s}$ of the muon track. ]{
The  $2p_{\mu}^{CM}/\sqrt{s}$ of the tag track with all other selection
criteria
applied for about 15\% of the \babar\ data sample. Contamination from
di-muon and Bhabha events, which peak at 2P/$\sqrt{s}$=1.0, are negligible. 
The points represent
the data, the empty histogram represents the
3 prong $\tau$ decays, the light shaded
histogram represents the other $\tau^{+}\tau^{-}$ MC, the medium dark
histogram represent the $\mu^{+}\mu^{-}$ MC and the dark histogram
represents the $q\bar{q}$ MC. 
  \label{fig:trkeff4:TagEPEP}}
\end{figure}

\begin{figure}
\centering
\includegraphics[width=0.45\textwidth]{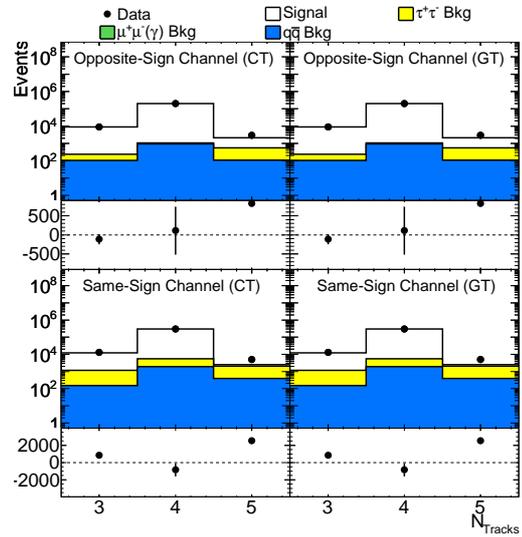}
\caption{ The track multiplicity in events that have been selected with
  the same-sign or opposite-sign selection presented using the CT and GT
  definitions of the fourth track with all
criteria
applied for about 15\% of the \babar\ data sample.  The points represent the data; the contributions
  from different backgrounds are shown in the histograms.
  \label{Tau31:multi}}
\end{figure}

After the same-sign and opposite-sign events have been selected,
fourth track candidates are selected, which are required to have the appropriate charge to come from a
$\tau$ pair event and satisfy the track definitions being studied.
Figure \ref{Tau31:multi} shows the multiplicity of the selected
same-sign and opposite-sign events for the CT and GT definitions.
Once the fourth track candidates have been selected, the tracking
efficiency is determined by using Eq. \ref{Tau31eff}.  The
difference in the tracking efficiency between data and MC is
defined using Eq. \ref{Tau31Delta}.

\begin{equation}
\Delta = 1- \frac{\epsilon_{MC}}{\epsilon_{data}}.
\label{Tau31Delta}
\end{equation}

%\Delta = 1- \frac{\epsilon_{MC}A}{\epsilon_{data}A}.

\noindent Similarly, the charge asymmetry of the tracking efficiency is defined using
Eq. \ref{Tau31ChAsym}.

\begin{equation}
a_{\pm} =\frac{\epsilon_{+} -  \epsilon_{-}}{\epsilon_{+} + \epsilon_{-}}
\label{Tau31ChAsym}.
\end{equation}

%a_{\pm} =\frac{\epsilon_{+}A -  \epsilon_{-}A}{\epsilon_{+}A + \epsilon_{-}A}
  
where the efficiency measurements in Eq.~\ref{Tau31Delta} and Eq.~\ref{Tau31ChAsym} also include the detector acceptance.

Monte Carlo studies indicate that the
backgrounds that could potentially bias the determination of the relative tracking efficiency
and the charge asymmetry are: events with two primary tracks from the $e^{+}e^{-}$
collision and a photon that converts into an electron pair; $q\bar{q}$ and
$\tau$ pair events with six
tracks; and  $\tau^-\to\pi^-K_{s}^{0}\nu_{\tau}$ where the $K_{s}^{0}$
decays into a $\pi^{-}\pi^{+}$ pair with a vertex that deviates
significantly from the primary vertex. For the background
events with conversions and  $K_{s}^{0}$, the reconstruction
efficiency could differ from that of tracks originating from the
interaction point of the $e^{+}e^{-}$ collision. The largest source of
conversions comes from hadronic $\tau$ decays with one charged track
and 1 or more neutral particles. This includes
$\tau^{\mp}\to\rho^{\mp}\nu_{\tau}$ and $\tau^{\mp}\to h^{\pm}\pi^{0}\pi^{0}\nu_{\tau}$ (h = $\pi$ or K),
which have branching fractions of $(25.51\pm0.09)\%$ and
$(9.51 \pm0.11 )\%$~\cite{pdg} respectively.
The contribution from 
the $\tau$ decays with a $K_{s}^{0}$ is small due to the
suppression by the selection cuts and the branching fractions.
The largest background from events with six
tracks originating from the $e^{+}e^{-}$ collision is from
$\tau^{\pm}\to\mu^{\pm}\nu_{\mu}\nu_{\tau}$, $\tau^{\mp}\to
h^{\mp}h^{\mp}h^{\mp}h^{\pm}h^{\pm}\ge0$ neutrals
$\nu_{\tau}(excluding \KS)$ events which have a branching fraction of 
$( 0.102\pm 0.004 )\%$. The contamination from $\tau$ pair events with
six tracks  is $\ll0.1\%$ for both the same-sign and opposite-sign channels.

\subsection{Systematic Uncertainties}

The primary systematic uncertainties in measuring the tracking
efficiency and charge asymmetry
arise due to mis-modeling of background contamination,
which can bias the tracking efficiency due to fake
tracks. 
%events where multiple fourth tracks are identified
The largest background comes from events with two tracks and a photon
that converts in the detector material producing an $e^{+}e^{-}$
pair. This background includes contributions from
$\tau$ pair events, radiative
di-muon,
Bhabha events, and two-photon events.

We estimate the effect of background mis-modeling on the efficiency measurement
using control samples selected to be enriched in photon conversion backgrounds.
The control samples are selected using the
standard selection, minus the vertex requirements, the 
loose electron rejection using PID, and the same-sign and opposite sign
invariant mass cuts. Instead of these we apply a tight electron PID selection
to two oppositely charged tracks. The invariant mass of the
two oppositely charged tracks is required to be less than 0.1 GeV using an
electron mass hypothesis.  The agreement between the data and MC for the
selection efficiency of this control sample is taken as the uncertainty
on the modeling of conversions.  This is propagated to the  $\Delta$ and
the charge asymmetry measurements using the measured rates. 
Note that this systematic error includes
both contributions from the mis-modeling of the conversions, and
contributions from
backgrounds that are not included in the MC simulation.

To assess the impact of potentially different track multiplicity from $q\bar{q}$
backgrounds and the small contribution from $\tau$ decays with a
$K_{S}^{0}$, the efficiency difference $\Delta$ and the charge asymmetry are calculated
without subtracting these backgrounds.  The difference between these and the nominal
values computed after background subtraction 
is conservatively taken as the systematic uncertainty.
%The uncertainty from
%the modeling of $\tau$ pair events with six or more tracks is negligible.

To account for possible differences in the rate of fake tracks, a systematic uncertainty based on the
difference between $\Delta$
and the charge asymmetry calculated with
$(\epsilon A)$ and 
\begin{equation}
\epsilon^{\prime}\times  A = \frac{N_4}{N_3 + N_4+N_5}
\label{Tau31effprime}
\end{equation}

\noindent is included, where $N_{5}$ is the number of events where
two candidate fourth tracks are found.

As a cross check on the systematic errors, we compute $\Delta$ and the
charge asymmetry, $a_{\pm}$, separately in
the same-sign and opposite-sign channels, and find these
to be consistent within statistical and background uncertainties.

%The Tau31 study provides a global measurement of the tracking
%efficiency convoluted with the geometric acceptance, the $\Delta$ and
%the charge asymmetry.

In general, tracks selected in an analysis will not have the same
kinematic distributions as the tracks in the Tau31
study. Therefore, when applying the efficiency results of the Tau31 study to an
analysis, an additional systematic
uncertainty is needed to account for the efficiency dependence on
track kinematics.  In the Tau31 analysis we do not estimate the dependence of
tracking efficiency on track density or track multiplicity.  That is done in the
ISR analysis presented in Section~\ref{sec:ISR}.
 
\begin{figure}
\centering
\includegraphics[width=0.45\textwidth]{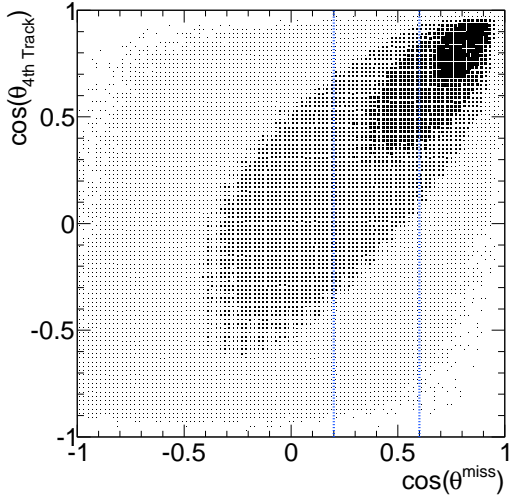}
\caption{The $cos(\theta_{4th\ Track })$ as a function of the $cos(\theta^{miss})$
for data events in Runs 1-6 selected with the same-sign
and opposite-sign selection criteria. The fourth track is
identified using the CT definition. The dotted lines indicate the
boundaries of the  $cos(\theta^{miss})$ regions selected for
determining the systematic uncertainty on
$\Delta$ and the charge asymmetry as a function  $cos(\theta_{4th\ Track })$.
\label{Fig:Tau31:ThetaSys}}
\end{figure}

%$cos(\theta)$ and $p_t$.
We quantify the kinematic variation in $\Delta$ and the charge asymmetry
by measuring them as a
function of fourth track polar angle $\theta$ and transverse momentum $p_t$.
Because of the three missing neutrinos in the event,
$\theta$ and $p_{t}$ of the fourth track cannot be exactly determined.
We therefore construct
estimators based on the trajectories
of the muon and two identified pions, and use them to define
different kinematic regions.
We define the $cos(\theta)$ estimator to be
\begin{equation}
\begin{split}
cos(\theta^{miss})= 
cos\left(\theta_{\pi_{1}\pi_{2}}\right)
\label{Tau31thetamiss},
\end{split}
\end{equation}

\noindent where the $\pi_{1}\pi_{2}$ system is defined as the vector sum of
the two identified pions. The correlation between the
$cos(\theta^{miss})$ estimator and the $cos(\theta_{4th\ Track})$ is
shown in Figure \ref{Fig:Tau31:ThetaSys}.
 
\begin{figure}
\centering
\includegraphics[width=0.45\textwidth]{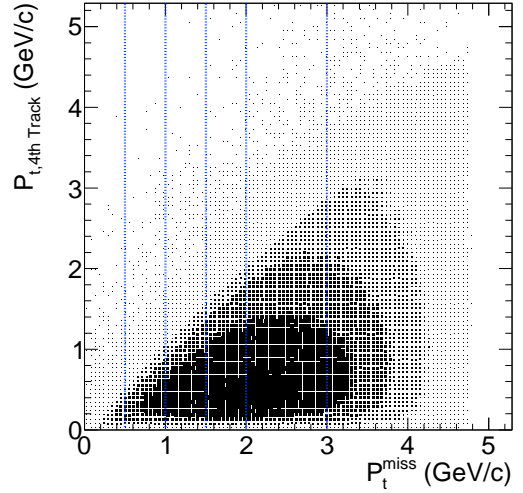}
\caption{ The $P_{t,4th\ Track }$ as a function of the $P_{t}^{miss}$
for data events in Runs 1-6 selected with the same-sign
and opposite-sign selection criteria. The fourth track is
identified using the CT definition.  The dotted lines indicate the
boundaries of the  $P_{t}^{miss}$ regions selected for
determining the systematic uncertainty on
$\Delta$ and the charge asymmetry as a function $P_{t,4th\ Track }$.
\label{Fig:Tau31:PtSys}}
\end{figure}

For $p_t$ we define the estimator as
\begin{equation}
p_{t}^{miss} =
\sqrt{ \left( \frac{\sqrt{s}}{2}-E_{\pi_{1}}-E_{\pi_{2}}  \right)^{2}-m_{\pi}^{2}}\times
  cos(\theta^{miss})
\label{Tau31ptmiss},
\end{equation}

\noindent where $\sqrt{s}$  is the beam energy, $E_{\pi_{1}}$ is
the energy of the  $i$th identified pion and $m_{\pi}$ is the mass
of a pion. The correlation between the
$p_{t}^{miss}$ and $P_{t,4th\ Track }$ is shown in Figure
  \ref{Fig:Tau31:PtSys}.

The systematic uncertainty on
$\Delta$ and the charge asymmetry as a function of the estimated $cos(\theta)$
and $p_t$ is defined as the
RMS$=\sqrt{ \sum_{i}^n\left( \Delta/a_{\pm}-\Delta_{i}/a_{\pm,i}\right)^2/\left( n-1 \right)}$, where $n$
is the number of regions, $\Delta/a_{\pm}$ is the average $\Delta$ 
or charge asymmetry as defined previously, and $\Delta_{i}/a_{\pm,i}$ is
the $\Delta$ or the charge asymmetry in the $i$th region selected with
the estimator.
The systematic uncertainty due to $P_{t}$ and $\theta$ dependence 
is quantified in Table \ref{tab:tau31sys}.

\begin{table}[h]
\begin{center}
\caption{Systematic uncertainties for $\Delta$ in $P_{t}$ and
  $\theta$. The $P_{t}$ and $\theta$ uncertainty in the  $\Upsilon$(2s) and
 $\Upsilon$(3s) runs are sensitive to the limited statistics.}
\begin{tabular}{lrr}
\toprule Data Period & $P_{t}$ Uncertainty & $\theta$ Uncertainty \\
\midrule
\multicolumn{3}{c}{GTL Track Definition}\\
Runs 1-6 & 0.20\% & 0.11\% \\
Run 7 - $\Upsilon$(2s) & 0.30\%  & 0.38\% \\
Run 7 - $\Upsilon$(3s) & 0.80\% & 0.42\% \\
\multicolumn{3}{c}{CT Track Definition}\\
Runs 1-6 & 0.10\% & 0.06\% \\
Run 7 - $\Upsilon$(2s) &0.65\%  & 0.28\% \\
Run 7 - $\Upsilon$(3s) &0.79\%  & 0.32\%\\
\bottomrule
\end{tabular}
\label{tab:tau31sys}
\end{center}
\end{table}

\subsection{Tau31 Results {\it vs.} Run Period}

\begin{figure}
\centering
\includegraphics[width=0.45\textwidth]{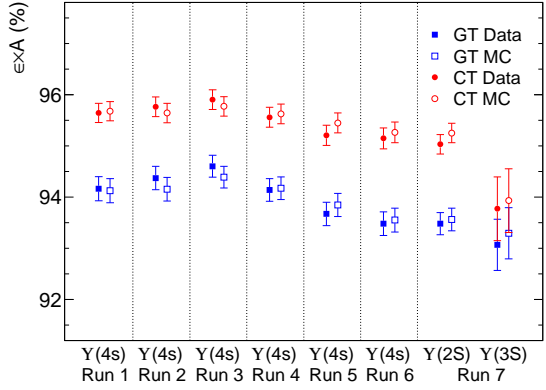}
\caption[The Tracking Efficiency as a Function of Run Number]{ The
  tracking efficiency as a function of run number for the GT
  definition (top)  and the CT definition
  (bottom). The data are represented by solid markers and the MC by open
  markers. The error bars represent the total uncertainty where the
  statistical and background systematic uncertainties have been added in
  quadrature. \label{fig:trkeff4:eff}}
\end{figure}

\begin{figure}
\centering
\includegraphics[width=0.45\textwidth]{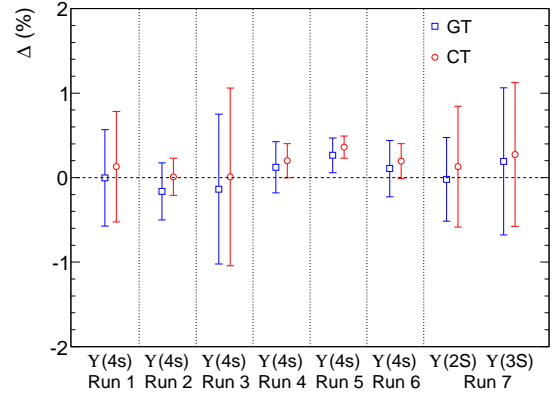}
\caption[The Data-MC difference in tracking efficiency as a Function of
  run number]{ The data-MC difference in the tracking efficiency as a
  function of run number for the GT definition and
  the CT definition. The error bars represent the total uncertainty where
  the statistical and systematic uncertainties have been added in
  quadrature. \label{fig:trkeff4:Delta}}
\end{figure}

\begin{figure}
\centering
\includegraphics[width=0.45\textwidth,bb=12pt 18pt 540pt 552pt]{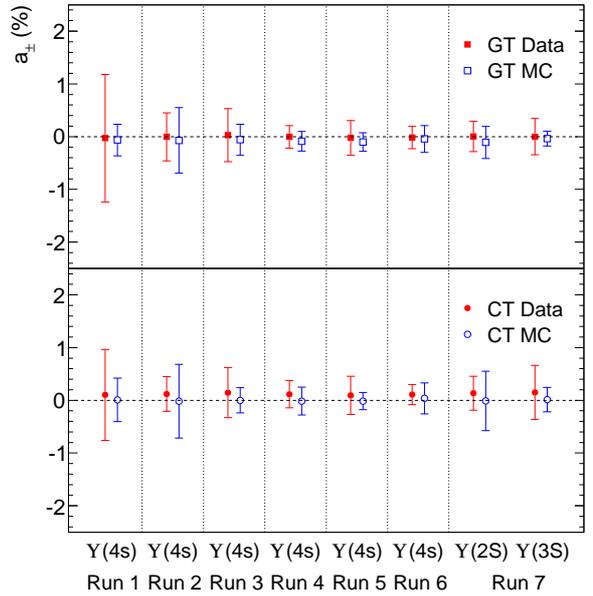}
\caption[Charge Asymmetry of the Tracking Efficiency as a Function of
  Run Number]{ The charge asymmetry of the tracking efficiency as a
  function of run number for the GT definition (top) and the CT definition (bottom).  The data are
  represented by the solid markers and the MC by the open
  markers. The error bars represent the total uncertainty where the
  statistical and systematic uncertainties have been added in
  quadrature.\label{fig:trkeff4:chasym}}
\end{figure}

Figure \ref{fig:trkeff4:eff} shows
the run-by-run tracking efficiency for the two track
definitions studied in this analysis: GT and CT.
The tracking efficiencies in the data and MC are found to be consistent
with each other. This can be seen in Figure \ref{fig:trkeff4:Delta}
which presents $\Delta$. The charge asymmetry
%of the tracking efficiency
can be seen in Figure \ref{fig:trkeff4:chasym}. The plots suggest that
there is no significant charge bias in the tracking efficiency.

The stability of the agreement between data and MC over the 7 run periods (as shown
in Figures \ref{fig:trkeff4:Delta} and \ref{fig:trkeff4:chasym})
demonstrates that the detector simulation, which is updated regularly,
accurately models the tracking performance of the detector as a function
of time. Because there is no significant time variation observed between
Runs 1 and 6 in $\Delta$ and in the charge asymmetry of the tracking
efficiencies, an average of Runs 1-6 for $\Delta$ and the charge
asymmetry of the tracking efficiencies is calculated.  These averages
are used to calculate the systematic uncertainty due to tracking
efficiency.  The systematic uncertainty per track for a given track definition is 

\begin{equation}
\Sigma_{Tracking}^{Tau31}=\frac{\sigma_{\Delta CT/GT}}{1-\Delta_{CT/GT}}
\label{Tau31uncert}
\end{equation}

\noindent  where $\sigma_{\Delta CT/GT}$ is the total uncertainty on $\Delta$ for
the given track definition. These results are the primary source of systematic 
uncertainty in track reconstruction efficiency in \babar.

\section{\bf Tracking efficiency using the ISR channel $\pipi\pipi\gamma_{ISR}$}
\label{sec:ISR}

A complementary approach to the Tau31 method is to study the tracking efficiency
using processes
such as  $\ep\en\to\pipi\gamma_{ISR}$ and $\ep\en\to\pipi\pipi\gamma_{ISR}$,
where a high energetic photon $\gamma_{ISR}$ is emitted from an initial lepton. This final state
provides a clean event sample, covering a wide range of momenta and polar angles of the tracks. In this section,
we describe one such measurement involving four pions in the final state along with the ISR photon. 
The Tau31 method has a higher statistical accuracy, allowing the explicit study of time
dependent effects.  By contrast, since no neutrinos are present in the final state, the ISR
events allow a more precise estimate of the missing track parameters than the Tau31 method.
In addition, the track density for ISR events is higher compared to the events in the Tau31 study,
corresponding to different~\babar\ physics channels.
The high track density in combination
with the precise track parameter prediction allows studying the
track overlap effects in tracking efficiency.

To study tracking efficiency with ISR, we use two event samples: one in which all 4
charged particles are reconstructed (4-track), and one in which only 3 charged particles
are found (3-track). Using energy and momentum conservation in a kinematic fit,
we can accurately predict the direction and momentum of the missing track in the 3-track sample.
By calculating the ratio of the number of lost tracks $N_{lost\hspace{0.05cm}tracks}$
to the number of measured tracks $N_{detected\hspace{0.05cm}tracks}$, we obtain the
tracking inefficiency, $\eta$, defined in equation (\ref{ineff}), and the tracking efficiency,
$\epsilon$, according to equation (\ref{eff}). Both can be measured as a function of the
kinematic properties of the missing track.

\begin{equation}
\eta= \frac{N_{lost\hspace{0.05cm}tracks}}{N_{detected\hspace{0.05cm}tracks} + N_{lost\hspace{0.05cm}tracks}}
\label{ineff}
\end{equation}

\begin{equation}
\epsilon=1-\eta
\label{eff}
\end{equation}

\subsection{ISR Event Selection}                                                                                                                                                   
For the ISR efficiency measurement we require that the tracks have a
polar angle inside the detector acceptance ($ -0.82 < \cos \theta_{ch} < 0.92$), and
that the transverse
distance of closest approach of the track to the event vertex (or nominal interaction point if
no primary event vertex is found) be smaller than 1.5\cm, and be within 2.5\cm in the beam direction.
Tracks with less than 100\mevc transverse momentum are
rejected. The ISR photon is restricted to the polar angular range inside the
EMC acceptance ($0.5\rad<\theta_{\gamma_{ISR}}<2.4\rad$), and a minimum photon energy
of $E_{ISR}>3\gev$ is required.  Either 3 or 4 selected tracks are required in the event.

%The ISR event selection is optimized to remove background events.
In order to suppress radiative Bhabha events, we reject events
where the two most energetic tracks pass a loose electron PID selection.
This also removes most $\g\g$ events
with an additional high energetic photon ($E_{\g, cm}>$ 4 \gev) in opposite direction
to the ISR photon candidate. We require the minimum angle between
the charged tracks and the ISR photon to be larger than $1.0\rad$, which
rejects a large fraction of
$\epem\rightarrow\qqbar$, $(\q=\u, \d, \s)$ 
and
$\epem\rightarrow\taup\taum$
event  backgrounds.
Events with one or two tracks with PID consistent
with being a $K^\pm$ in the 3-track or the 4-track sample are
rejected, respectively.  Finally, we require the $4\pi$ invariant mass
to be in the range of $1.2\gevcc<M_{4\pi}<2.4\gevcc$, where we expect 
a high signal to noise ratio.

Backgrounds from $\epem\rightarrow\qqbar$  $(\q=\u, \d, \s)$ are simulated
with JETSET~\cite{jetset}, while $\epem\to\taup\taum$ backgrounds are simulated using KORALB~\cite{koralb}.
The ISR-channels
are simulated with the AFKQED~\cite{afkqed} generator, based on an early version of
PHOKHARA~\cite{phok}. The MC samples are normalized according to the luminosity observed in the data.

\subsection{ISR Kinematic Fit}

Selected events are subjected to a kinematic fit assuming the $\pip\pim\pip\pim\gamma_{ISR}$
signal hypothesis $\chisq_{4\pi}$, as well as the $\Kp\Km\pip\pim\gamma_{ISR}$
background hypothesis $\chisq_{2K2\pi}$. The fit in the 4-track (3-track) sample uses
the four (three) tracks, the ISR photon and the kinematic information of incoming electron
and positron. Energy and momentum conservation leads to four (one) constraints, or a
4C-fit (1C-fit), respectively.

\begin{figure}[h!]
\centerline{\includegraphics[clip, width=6.cm]{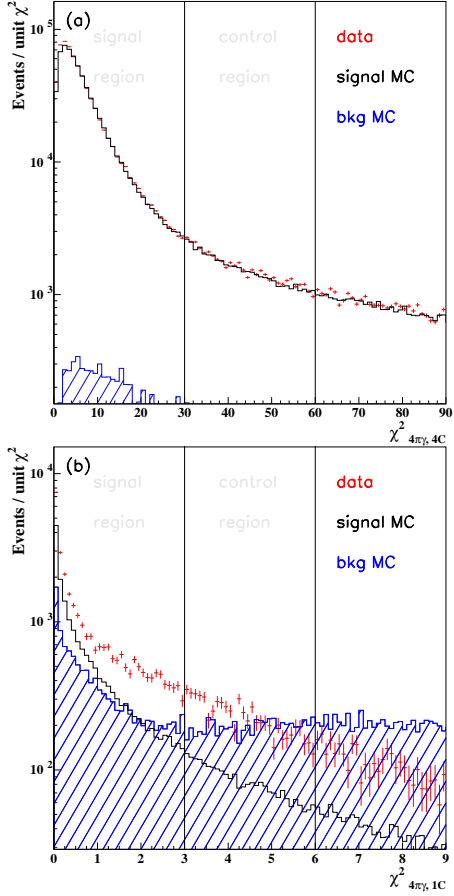}}
\vspace{-0.3cm}
\caption{(a): $\chisq_{4\pi}$ distribution for 4-track sample (4C). Data with subtracted
background (red points), signal MC (black histogram) and the sum of background MC
channels (blue histogram). (b): Corresponding $\chisq_{4\pi}$ distributions for the
3-track sample (1C). The signal and control regions are indicated with vertical lines, with 
the region in the extreme left being the signal region and the area in the middle being the control region.}
\label{chi2}
\end{figure}

The resulting $\chisq$ distributions are shown in Fig.~\ref{chi2}.
The $\chisq$ distributions are broader than expected, partly
due to detector resolution effects, but mostly because additional ISR photons are not
included into the kinematic fit hypothesis.
In Fig.~\ref{chi2}(a) the 4-track sample shows a good agreement
between the data and MC in the presence of negligible background.
%The kinematic fit $\chisq$ is used to make the final selection in the 4-track
%sample $\chisq_{4\pi,4C} < 30$.

In Fig.~\ref{chi2}(b) the corresponding $\chisq$ distributions are shown for the
3-track sample.  Here, we also require the predicted polar angle for
the missing track be in the detector acceptance  ($ -0.82 < \cos \theta_{ch} < 0.92$) region.
The relative amount
of background is much larger in this sample, since the kinematic closure that suppresses a
lot of background in the 4-track sample is weaker with only one constraint.
The visible difference between the number of events after background subtraction (red)
and signal MC (black) suggests that
more $\pi$ tracks are lost in data than are described by MC.

\begin{figure}[h]
\begin{center}
\includegraphics[clip, width=0.5\textwidth]{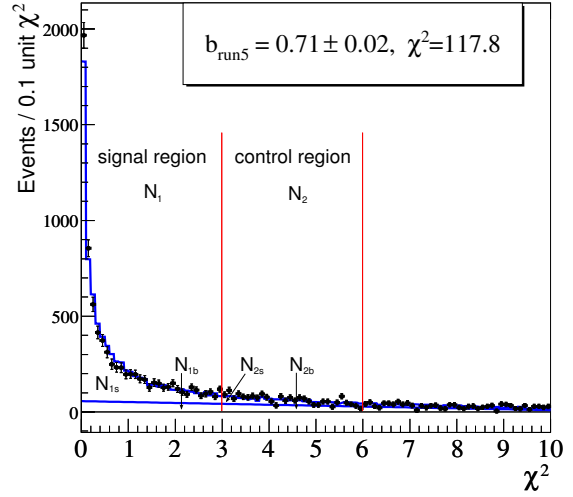}
\end{center}
\vspace{-0.3cm}
\caption{Fit result for sideband parameter $b$ using Run 5 fitting
signal MC (blue histogram) and a linear background (blue line)
to data (black points). Also indicated are the number of signal $N_{1s/2s}$ and background $N_{1b/2b}$ events in the signal and control region.}
\label{sideband}
\end{figure}

Fig.~\ref{chi2}(b) also shows a difference in the shape of the $\chisq$ distributions
between the data and MC. The plateau in the background MC at large $\chisq$
suggests that all backgrounds are not subtracted from data.
Therefore we perform an additional background subtraction based on data sidebands.
%\begin{comment}                                                                                                                                                                               
The idea of the sideband subtraction is illustrated in
Fig.~\ref{sideband}, which plots the 3-track $\chisq$ distribution for a subset
of the \babar\ data.  We define a signal
region enriched in signal events, which contains $N_1$ events.  The control region,
which has substantial background contributions, contains $N_2$
events. Let  $N_{1s}$ ($N_{1b}$) be the number of signal (background) events
in the signal region, and $N_{2s}$ ($N_{2b}$) the corresponding numbers for
the control region. Assuming one knows the ratios,

\begin{equation}
a=\frac{N_{2s}}{N_{1s}}\hspace{0.3cm} and
\hspace{0.3cm}b=\frac{N_{2b}}{N_{1b}}
\label{aEquation}
\end{equation}

\noindent the number of signal events can then be calculated according to the following equation:

\begin{equation}
N_{1s}=\frac{b\cdot N_1 - N_2}{a-b}
\label{N1sEquation}
\end{equation}

We define signal and control
regions in the 4-track sample as $\chisq_{4\pi,4C} < 30$ and
$30<\chisq_{4\pi,4C} < 60$ respectively.
The corresponding regions in the 3-track sample
are chosen so that the ratio of events in the signal to
control region is the same as in the 4-track sample, resulting in
$\chisq_{4\pi,1C} < 3$ and $3 < \chisq_{4\pi,1C} < 6$ respectively. The
ratio $a$ is determined using signal MC. In order to obtain $b$, we
assume any difference in tracking inefficiency between data and MC
does not depend on
$\chisq_{4\pi}$. Therefore we performed a fit of the
difference between data and MC using a linear Probability Density
Function (PDF),  allowing a scale-factor for MC.

The result of the fit is shown in Fig.~\ref{sideband}. Small
discrepancies at low $\chisq$ indicate that there is still some
background present. The remaining difference in the $\chisq$ distribution is consistent within the 
uncertainty of the cross section of the peaking background contributions that have been subtracted.
After subtracting the additional background using equation \ref{N1sEquation}, the inefficiency difference
between data and MC $\Delta\eta = \eta_{data} - \eta_{MC}$ is determined to be

\begin{equation}
\Delta\eta =  (0.75 \pm 0.05_{stat} \pm 0.34_{syst})\%.
\label{ineff_diff}
\end{equation}

The systematic uncertainty on $\Delta \eta$ is dominated by the uncertainty of the cross section of the subtracted individual background contributions in the 3-track sample. Most of these cross sections have been measured in previous \babar\ analyses ~\cite{3pi, 6pi, 5pi, 2k2pi_new, 2kpi}. The normalization of the additional contributions of continuum and $\ep\en\to\tau\tau$ backgrounds have been verified with specific kinematic distributions.
Note that this result is not directly comparable to the Tau31 efficiency result, as that was
calculated using an isolation requirement between the tracks.  The effect of track overlaps
is discussed in the next section.

\subsection{ISR Efficiency Kinematic Dependence}

\begin{figure}[h]
\centerline{\includegraphics[clip, width=6.cm]{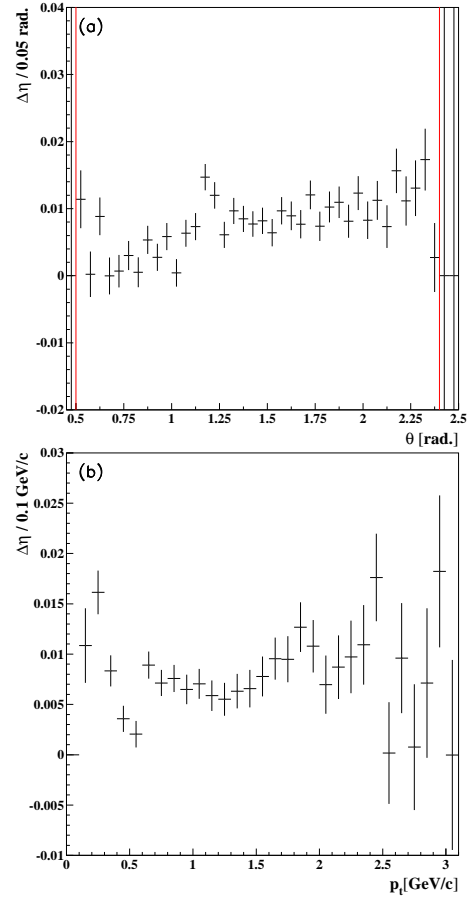}}
\vspace{-0.3cm}
\caption{Relative data-MC difference of tracking inefficiency vs. the polar angle
of the track $\theta$ (a) and vs. the transverse momentum $p_t$ (b). Red lines
indicate the detection region used to determine the average inefficiency.}
\label{tkr_diff_theta}
\end{figure}

In Fig.~\ref{tkr_diff_theta} the dependence of $\Delta\eta$ on the polar
angle $\theta$ (a) and the transverse momentum $p_t$ (b) of the missing track is presented. The
dependence on $p_t$ is flat within the uncertainties of 0.4\%. A slight
dependence on the polar angle is visible with almost no difference between
data and MC in the forward region at small polar angles and a difference of
approximately 1$\%$ in the central and backward region. Due to the beam energy
asymmetry at~\babar, high energy photons are preferably emitted in the forward
direction at small polar angles. In ISR events, the hadronic system is emitted
back-to-back to the ISR photon. The energy of the photon is correlated with the
opening angle of the cone of the hadronic system. This correlation leads to an
increasing track overlap probability in the backward region of the detector,
which is not perfectly modeled by MC as shown in the following.

One source of tracking inefficiency is when two tracks overlap in the detector,
causing sensor signals from one or both to be lost or distorted, and creating hit
patterns that can be hard for the track finding algorithms to distinguish.
\babar~tracking
inefficiency is most affected by overlaps in azimuth, as the DCH largely projects
out track polar angle.  Due to magnetic bending, tracks with the same charge are
more likely to overlap in azimuth than tracks with opposite charge.  Furthermore,
the overlap between tracks with opposite charge depends in an asymmetric way on the
azimuthal angle between them.  These effects are shown schematically in
Fig.~\ref{overlap_comic}. To study the dependence of tracking efficiency on overlap,
we define variables sensitive to the charge-dependent two-track azimuthal separation:

\begin{figure}[h]
\centerline{\includegraphics[clip, width=7.cm]{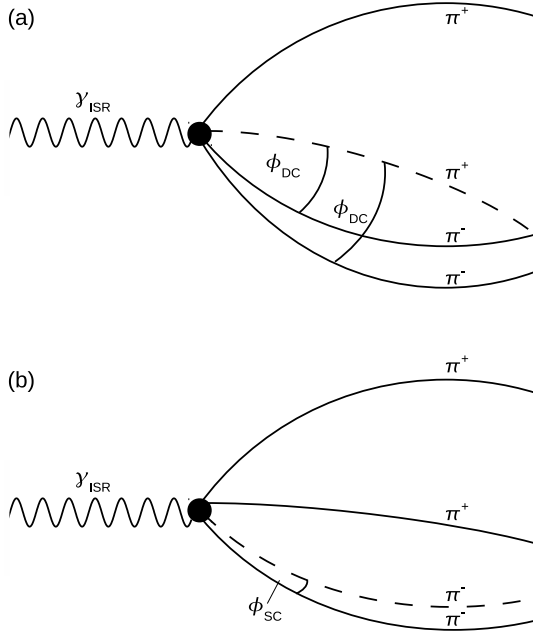}}
\vspace{-0.3cm}
\caption{(a): Charge oriented azimuthal angle between missing pion and detected
pions with different charge $\Delta\phi_{DC}=\phi(\pip)-\phi(\pim)$ (2 entries per
event); (b) absolute value of azimuthal angle between missing pion and detected pion
with same charge $\Delta\phi_{SC} = |\phi(\pipm)-\phi(\pipm)|$.}
\label{overlap_comic}
\end{figure}

\begin{figure}[h]
\centerline{\includegraphics[clip, width=8.cm]{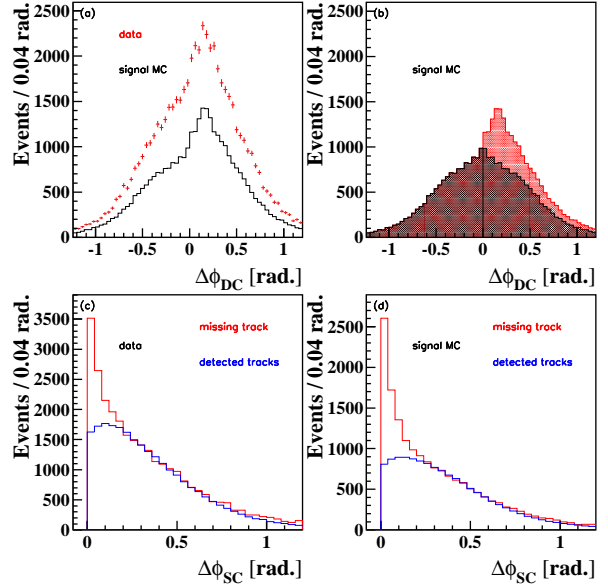}}
\vspace{-0.3cm}
\caption{(a): Angle between missing pion and detected pions with different charge
for the data (red points) and signal simulation normalized to the luminosity (black histogram). Two entries per event.
(b): Illustration of the cleaning procedure. Tracks lost due to DC overlap (red) and
due to other effects (black). (c): Angle between missing pion and detected pions with same charge (red) and the angle 
between two detected pions (blue) normalized to the same number of events in the region $0.3\rad<\Delta\Phi<0.8\rad$. (d): Same as (c), but
for signal MC.}
\label{ov_dc_all}
\end{figure}

The effect of track loss due to overlapping tracks with different charge (DC) is
visible in the distribution of the charge oriented azimuthal angle difference
between the lost track and the reconstructed track with different charge
$\Delta\phi_{DC} = \phi(\pip)-\phi(\pim)$. Since in our study there are always
two pions with the different charge of the missing pion, two angles are obtained
for each event. In Fig.~\ref{ov_dc_all}(a) the $\Delta\phi_{DC}$ distribution
is plotted for data and MC.
%Without track overlap a symmetric distribution around 0 is expected. 
The asymmetric
distribution shows that the DC tracking inefficiency peaks at small positive values.

The number of tracks lost due to DC track overlap is estimated by subtracting the
negative half of this distribution from the positive, as
illustrated for MC in Fig.~\ref{ov_dc_all}(b). The inefficiency is corrected as indicated
in equation ~\ref{ineffcor}, leading to a correction for overlapping tracks with different
charge DC of $0.41\%$.

\begin{equation}
\eta'= \frac{N_{lost\hspace{0.05cm}tracks} - N_{overlapping\hspace{0.05cm}tracks}}{N_{tracks}}
\label{ineffcor}
\end{equation}

We describe the same charge (SC) track overlap inefficiency in terms of $\Delta\phi_{SC} = |\phi(\pipm)-\phi(\pipm)|$, as 
illustrated in Fig.~\ref{overlap_comic} (b): the angle between the lost track and the reconstructed track with
the same charge.  For data in Fig.~\ref{ov_dc_all}(c) the angle between
lost track and reconstructed track with the same charge in the 3-track
sample is plotted in red. The blue histogram shows the same distribution
for the two detected tracks.  The distribution with one lost track is
the superposition of the distribution due to detection inefficiency and
a peaking distribution at small $\Delta\phi_{SC}$ due to track overlap
losses. The distribution due to usual detection inefficiency has the
same $\Delta\phi_{SC}$ dependence as the distribution of the two
measured tracks.  The number of tracks lost due to track overlap can be
estimated by scaling down the distribution of the measured tracks until
the tails of the distribution match with the distribution including one
missing pion. The difference at small $\Delta\phi_{SC}$ is a good
estimate for the number of tracks lost due to track overlap. The
corresponding distributions for MC are displayed in
Fig.~\ref{ov_dc_all}(d). The effect of SC tracks overlap is well modeled in MC.

%The effect of overlapping SC tracks
%on the track reconstruction efficiency is well modeled by MC.

\subsection{ISR Efficiency Summary}

To summarize, the difference in tracking inefficiency per track
including track overlap is determined from ISR events to be:

\begin{equation}
  \Delta\eta = (0.75 \pm 0.05_{stat} \pm 0.34_{syst})\%
\label{EtaEqn}
\end{equation}

%\noindent To compare this with the Tau31 measurement, we apply the same track isolation
%cuts as were used in that analysis.
%This leads to the following result, which agrees within the uncertainties with the Tau31
%study:

\noindent Because of the track isolation requirement applied in the Tau31 selection, the 
different track multiplicity, and the different event topology, the ISR study includes a 
significantly higher track overlap probability and thus the value in equation \ref{EtaEqn} 
is not directly comparable with the Tau31 result discussed in Section~\ref{sec:tau31eff}.
To make a comparison, we quantify the effect due to track overlap by studying the distributions 
of the azimuthal angular difference between same charged tracks and oppositely charged tracks.
Taking this effect into account, we measure an efficiency difference between data and simulation of:

\begin{equation}
  \Delta\eta' =  (0.38 \pm 0.05_{stat} \pm 0.39_{syst})\%
\label{EtaPrimeEqn}
\end{equation}

This result is consistent with the Tau31 efficiency difference within the uncertainties.
Depending on the event multiplicity and kinematics, \babar\ analyses may need
the inefficiency with or without track overlap effects.

\section{\bf Tracking Charge Asymmetry}
\label{TrackingChargeAsymmetry}

Since a main objective of the \babar\ experiment is to measure CP
violation, it is vital to understand
and measure any possible charge asymmetry in the track reconstruction. For
instance, a promising mode for
searching for CP violation in charm decays is $D^\pm \to K^+K^-\pi^\pm$.
An asymmetry in the reconstruction efficiency for the $\pi^\pm$ would
bias the CP result.
Because the signal in these
decays has a statistical uncertainty of $\sim 0.25$\%, a
comparable control of the tracking efficiency asymmetry is needed.

We define the charged pion tracking asymmetry as
\begin{equation}
a(p_{Lab}) \equiv \frac{\epsilon(p_{\pi^+}) - \epsilon(p_{\pi^-})}
{\epsilon(p_{\pi^+}) + \epsilon(p_{\pi^-}) }
\label{eq:asym}
\end{equation}

\noindent where $p_{Lab}$ indicates that momenta are in the lab frame, and
$p_{\pi^{+}}$ ($p_{\pi^{-}}$) refers to the momentum of
the positively (negatively) charged pion.

\begin{figure}
  \centering
  \includegraphics[width=0.49\textwidth]{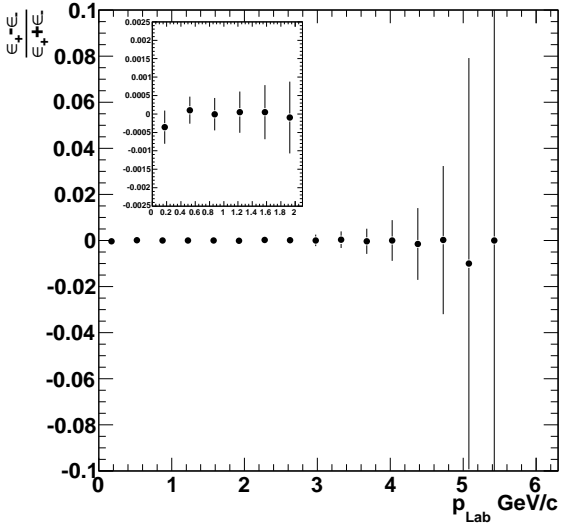}
  \caption{Tracking asymmetry from MC as a function of charged pion
    momentum in the laboratory frame. The inset plot shows the asymmetry up to 2 GeV.}
  \label{fig:MC_eff}
\end{figure}

We illustrate our expectations in this regard using MC.
Figure~\ref{fig:MC_eff} shows the pion tracking efficiency asymmetry
derived from MC using generator information for pion tracks in $D^\pm \to
K^+K^-\pi^\pm$ decays. The average asymmetry for MC in this mode is found to be
$a(p_{Lab}) = (-6 \pm 23) \times 10^{-5}$, consistent with zero within the
uncertainties, and without any significant momentum dependence.

\begin{figure}
\centering
\includegraphics[width=0.49\textwidth]{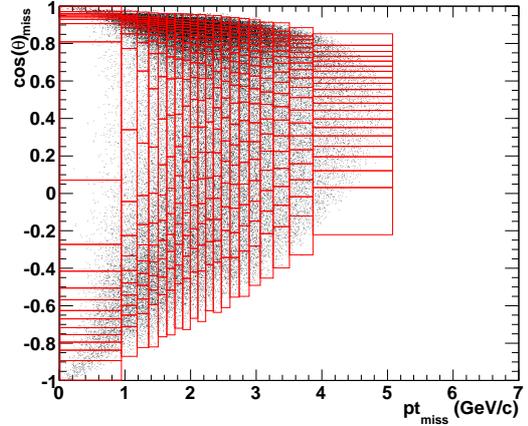}
  \caption{Distribution of the observed 2-hadron events for the full
    \babar~ data sample. Bin boundaries were chosen to obtain the same number
    of events (215) in each bin.}
  \label{fig:2D}
\end{figure}

Two different methods are used to determine the pion track efficiency
asymmetry directly from data. The more precise technique relies on
Tau31 events. We work directly in the observed variables and use the
ratios of the numbers of two-hadron decays to three-hadron decays to determine the pion
inefficiency. Instead of fitting distributions of 2- and 3-hadron
decays, we recognize that the (fewer) 2-hadron events that arise from
tracking inefficiency can be easily modeled directly from the 3-hadron
events. In practice this is done by weighting every 3-hadron event by
the ratio $(1-\epsilon)/\epsilon$, where $\epsilon$ is the track
efficiency of the observed third track. For both 3-hadron as well as
2-hadron events we select only events from the $\rho$-decay channels 
$\tau^{-}\rightarrow\rho^{0}h^{-}\nu_{\tau}$, according to the selection
criteria described in section \ref{sec:tau31eff} since the inclusive 
$\tau^{-}\rightarrow\pi^{-}\pi^{-} h^+\nu_{\tau}$ has more significant
backgrounds, specifically with contamination from electrons. The total
number of 2-prong (3-prong) events in the sample is 86,092
(1,365,900). The distribution of events in the observed variables, 
$pt_{miss}$ and $\cos(\theta)_{miss}$, is shown in Figure~\ref{fig:2D}. 
The observed variables are determined from the 2-prong momenta:
\begin{equation}
  \vec p(\pi\pi) \equiv \vec p(\pi^+) + \vec p(\pi^-)
\end{equation}
such that 
\begin{equation}
   \cos(\theta)_{miss} = \frac{p_z(\pi\pi)}{p(\pi\pi)}
\end{equation}
and
\begin{equation}
        pt_{miss} = p_T(\pi\pi).
\end{equation}

%Unfolding the distribution in these two variables one can then
%determine the true distribution in pion track angle and missing $p_T$.
%Due to uncertainties in unfolding, 
%require that two oppositely-charged tracks be consistent with being from a $\rho^0$ decay.

%We apply several criteria on data events to reduce them to the final
%set for the efficiency measurement. The cuts which reduce data the
%most are the requirement that there be an isolated muon on the tag
%side, the signal side be multi-prong, a $dE/dx$ requirement on the pion
%track, there be no neutral clusters, and an electron veto on the pions
%in the $\rho^0$ decay. All of these cuts except the last two also
%improve signal to background significantly. 

In order to fit these event distributions, one must also account for
backgrounds.  The 2-hadron events of interest include approximately
7\%\ background events. Chief among these are events from photon (5\%) and
$\pi^0$ (1\%) conversions in 1-hadron decays of the tau, where the
1-prong track from the tau is combined with a track from the photon or
$\pi^0$ and identified as a 2-prong event. Inelastic nuclear
interactions due to tracks passing through detector material and other
backgrounds are small in comparison. The backgrounds are split into
``photon'' and ``other'' components and the overall normalization of
each distribution is a parameter in a binned $\chi^{2}$ fit. Another
large background contribution to 2-hadron events (whose normalization is
a parameter) is acceptance loss events due to the third track being lost
in the direction of the beam. PDFs are obtained from MC as normalized
histograms in the observed variables of the various backgrounds; events
in these have been re-weighted to account for inadequacies in the MC
3-body Dalitz distributions by matching the 3-body mass distribution as
well as both the 2-body mass distributions to those in data. The
tracking efficiency asymmetry fit is a binned $\chi^2$-fit with binning
as shown in Fig.~\ref{fig:2D}.

The significant parameters in the fit describe the tracking efficiency
and the asymmetry as a function of the lab momentum. The tracking
efficiency is parameterized with the following phenomenological formula:

\begin{equation}
\epsilon(p_{Lab}) = 1 - A_{0}e^{\frac{p_{Lab} - p_{0}}{\tau_{0}}} - B_{0}e^{\frac{p_{Lab} - p_{1}}{\tau_{1}}}
\end{equation}

\noindent where the parameters are $A_{0}$, $B_{0}$, $p_{0}$, $p_{1}$, $\tau_{0}$, and $\tau_{1}$, 
in addition to parameters which measure the asymmetry in bins of lab momentum. 
Finally, it should be mentioned that we account for differences in the 3-hadron 
distributions of $m^{2}_{12}$ versus $m^{2}_{23}$ ($1,2,3$ denote the particles 
in the 3-prong tau decay) in data and MC by weighting 3-hadron events according 
to the data/MC $m^{2}_{12}$ , $m^{2}_{23}$ distribution ratio. 
The fit to our data is good as evidenced by a $\chi^2/$NDF = 792/780, i.e., a 37\%\ probability.

\begin{figure}[htbp]
  \centering
  \includegraphics[width=0.49\textwidth]{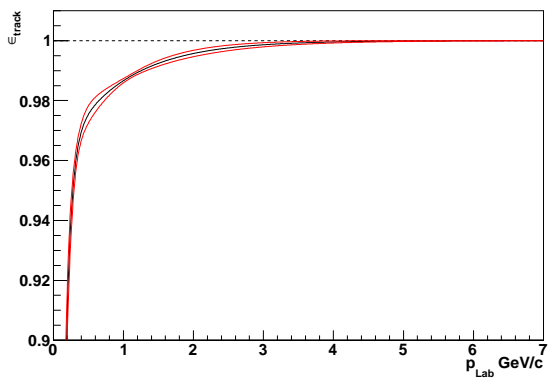}
  \caption{The tracking efficiency determined by the Tau31
    method as a function of charged pion momentum in the laboratory frame. The red envelope
    around the efficiency curve indicates 1$\sigma$ statistical error
    bands.} 
   
  \label{fig:tau31_results1}
\end{figure}

\begin{figure}[htbp]
  \centering
  \includegraphics[width=0.49\textwidth]{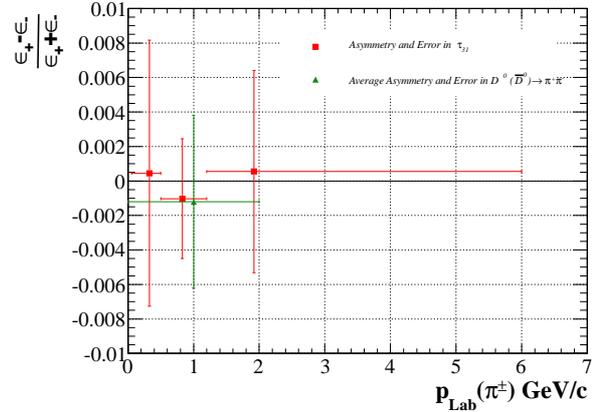}
  \caption{The tracking asymmetry determined by the Tau31
    method as a function of charged pion momentum in the laboratory frame. The average asymmetry
  over momenta 0-2 GeV/c determined from $D^0$ decays is also shown here 
  for comparison.}
  \label{fig:tau31_results2}
\end{figure}

Results from this procedure are shown in Figures~\ref{fig:tau31_results1} and \ref{fig:tau31_results2}.
We find the average charged pion tracking efficiency asymmetry to be 
$a(p_{Lab}) =  (0.10 \pm 0.26)\%$, in our momentum range of approximately 0-4 GeV/c,
consistent with zero. To account for systematic errors we re-fit the
data with the following variations. We force the acceptance loss and
background descriptions in the fit individually to be
charge-independent, and we reduce the number of background components by
combining some PDFs. We find the total systematic error to be 0.10\%.

Another technique we use to measure the charged track efficiency
asymmetry utilizes isotropy of spinless-two-body decays. In this method
we study the $D^0\to \pi^+\pi^-$ and $\overline{D}^0\to \pi^+\pi^-$ decays. We require that these decays not
be from B-meson decays (as these have larger backgrounds) and that they
be tagged as being from $D^{*\pm}$ decays to improve signal purity.
Also, in both cases we require that at least one pion have momentum
greater than 2 GeV$/c$ and assume that the tracking efficiency charge
asymmetry is zero for this pion. Therefore, any asymmetry in yields is
the result of tracking asymmetry of the lower momentum pion which is
reported below.

High purity samples of $D^{0}$ and $\overline{D}^{0}$ decays are
obtained using slow pions associated with the decay of $D^{*+}$ to tag
the flavor of the $D^{0}$ meson. A detailed description of the event
selection is described in the publication of $D^{0}-\bar{D}^{0}$ mixing
using the ratio of lifetimes for the decay of
$D^{0}\rightarrow\pi^+\pi^-$ \cite{BaBar:mixing}. Particle
identification is not applied to the selection of pion tracks, rather we
choose to remove reflections from the $K^-\pi^+$ decays of the $D^0$
using a cut on the reflected mass and we account for the remaining
contamination from the tails by studying their $\pi^+\pi^-$ mass
distributions and including a term with such a shape in our 1-D binned
$\pi^+\pi^-$ mass fit. Yields of $D^{0}$ decays where the higher
momentum track is either the $\pi^{+}$ or the $\pi^{-}$ are
separately determined and are used to determine the asymmetry. A similar
study is carried out using $\overline{D}^{0}$ decays,
and the combined charge asymmetry of the efficiency,
averaged over pion momenta from 0 to 2 GeV/c is found to be
$a(p_{Lab})  = (-0.12 \pm 0.50)\%$, consistent
with zero and the Tau31 method result, but not as precise as the Tau31 method.

\section{\bf Low $p_T$ tracking efficiency measurement}

The $\tau$ pair sample provides an estimate of tracking efficiency for
charged tracks with $p_T >$ 180 MeV/c only.  However, the detection of
low $p_T$ tracks ($p_T <$ 180 MeV/c) is important for tagged \Dz
analyses. \Dz tagging is performed through the \Dstarp\to\Dz\pip decay,
where the {\em soft pion} ($\pip_s$) is emitted with an energy just
over its rest mass in the $\Dstarp$ frame, and so typically has very low
$p_T$ in the lab frame.  \Dz tagging
is used in \CP violation, mixing, and many other precision analyses,
therefore a good understanding of the low $p_T$ tracking efficiency is required.

\begin{figure}[htb]
\begin{center}
\includegraphics[width=0.5\textwidth]{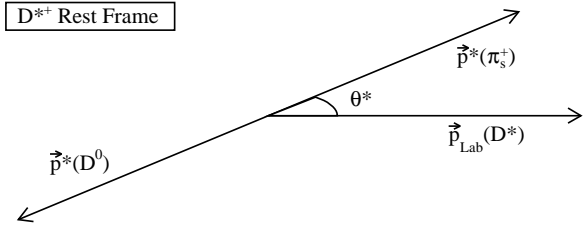}
\caption{Definition of slow pion helicity angle $\theta^*$.}
\label{fig:SoftPiHelicity}
\end{center}
\end{figure}

The low $p_T$ reconstruction efficiency
analysis is based on a previous analysis by the CLEO collaboration \cite{Menary:1992}.
CLEO demonstrated that the relative slow pion efficiency
can be measured as a
function of momentum using {\em helicity}
distributions.
The slow pion helicity angle $\theta^*$ is defined as the angle between
the slow pion momentum in the \Dstar rest frame and the \Dstar momentum
in the laboratory frame. This is illustrated in Fig.~\ref{fig:SoftPiHelicity}.

When a vector meson decays to a final state made of two pseudo-scalar
mesons, the distribution of the helicity angle is expected to be symmetrical and
can be described as~\cite{Brandenburg:1998ap,Gibbons:1997ag}
\begin{equation}
\frac{dN}{d\cos\theta^*} \propto (1+\alpha\cos^2\theta^*), \quad
1<\alpha<+\infty,
\label{eq:SoftPiCosTh}
\end{equation}

\noindent Furthermore, the cosine of the helicity angle is related to the slow pion momentum by:
\begin{equation}
p_{\pi_s} = \gamma(p^*_{\pi_s}\cos\theta^* - \beta E^*_{\pi_s}),
\label{eq:SoftPiMomentum}
\end{equation}
where $\beta$ and $\gamma$ are the \Dstar boost parameters.
Since $p^*$ and $E^*$ are known once the \Dstar momentum is known,
Eq.~\ref{eq:SoftPiMomentum} maps any asymmetry observed in Eq.~\ref{eq:SoftPiCosTh} to a
relative reconstruction inefficiency in a
specific part of the slow pion momentum spectrum.

\begin{figure}[h]
\centering
\includegraphics[width=0.45\textwidth]{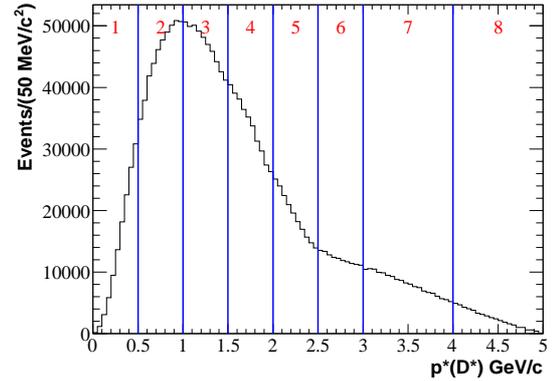}
\caption{\label{fig:SoftPiPstarBins}Distribution of $p^*(\Dstar)$ in the data sample.
On top of the figure the blue lines show the lower and upper limit of each bin
indicated by the red number.}
\end{figure}

We measure the $\cos\theta^*$ distribution in 8 bins of $p^*(\Dstar)$
spectrum as shown in Fig.~\ref{fig:SoftPiPstarBins}.
Since $p_{\pi}$ depends not just on the $\cos\theta^*$, but also on
$p^*(\Dstar)$, we perform an angular efficiency analysis in bins of
$p^*(\Dstar)$.  We then fit these $\cos\theta^*$ distrubions to a
function defined as the convolution of Eq.~\ref{eq:SoftPiCosTh} and the
efficiency function:

\begin{equation}
\epsilon(p) =
\begin{cases}
1 - \frac{1}{\delta(p-p_0) + 1}, & \text{ if } p > p_0\\0, & \text{ if }
  p \leq p_0.\
\end{cases}
\label{eq:SoftPiEff}
\end{equation}

%\begin{table}[h]
%\begin{center}
%\caption{$p^*(\Dstar)$ binning.}
%\begin{tabular}{cc}
%\toprule bin & range \\
%\midrule
% 1      & $p^*(\Dstar) \leq 0.5$\gevc\\
% 2      & $0.5 < p^*(\Dstar) \leq 1.0$\gevc\\
% 3      & $1.0 < p^*(\Dstar) \leq 1.5$\gevc\\
% 4      & $1.5 < p^*(\Dstar) \leq 2.0$\gevc\\
% 5      & $2.0 < p^*(\Dstar) \leq 2.5$\gevc\\
% 6      & $2.5 < p^*(\Dstar) \leq 3.0$\gevc\\
% 7      & $3.0 < p^*(\Dstar) \leq 4.0$\gevc\\
% 8      & $4.0 < p^*(\Dstar) \leq 5.0$\gevc\\
%\bottomrule
%\end{tabular}
%\label{tab:SoftPiPstarBins}
%\end{center}
%\end{table}

%where $p^*_{\pi_s}$, $E^*_{\pi_s}$ are the momentum and the energy of the
%slow pion evaluated in the \Dstar rest frame, respectively, while $\beta$ and
%$\gamma=\sqrt{1-\beta^2}$ are the boost parameters. Since $p^*_{\pi_s}$ and 
%$E^*_{\pi_s}$ are given once the \Dstar momentum is known, Eq.~\ref{eq:SoftPiMomentum} 
%maps any asymmetry observed in Eq.~\ref{eq:SoftPiCosTh} to a specific part of the slow pion momentum spectrum.                                                                                                                     

The goal of this analysis is to compare data and MC efficiencies to
get a systematic error from the relative difference between them:
\begin{equation}
\sigma_{syst} = \frac{ \int \epsilon_{data}(p)dp - \int
  \epsilon_{MC}(p)dp}{\int \epsilon_{data}(p)dp}.
\label{eq:SoftPiSyst}
\end{equation}

\noindent The limitations of this method are the effects that may be not correctly
described in the MC, such as final state interactions or radiative
losses.

%This analysis is based on a previous analysis by the CLEO collaboration,
%that has demonstrated that the relative slow pion efficiency as a 
%function of momentum can be measured using helicity distributions\cite{Menary:1992}.                                                                                                                                           

The analysis is done using 470\invfb of data recorded by the
\babar\ detector and about $4.2\times10^9$ generic MC events.
The decay chain \epem\to\Dstarp X, \Dstarp\to\Dz$\pip_s$,
\Dz\to\Km\pip\cite{conj} is reconstructed in both data and MC,
requiring particle identification for the kaon and the two vertices to be successfully
reconstructed.   The \Dz\to\Km\pip mode is chosen to provide a
clean sample of \Dstarp\to\Dz\pip decays with a high
branching fraction.
A control sample is reconstructed the same way by not
requiring the kaon identification. This sample is used for background subtraction.
The $p^*(\Dstar)$ spectrum has been compared between data and MC.
Differences are corrected for by weighting the MC sample which is then
normalized to data.
% The following steps are made on data and MC as well.     

\begin{figure}[htb]
\begin{center}
\includegraphics[width=0.45\textwidth]{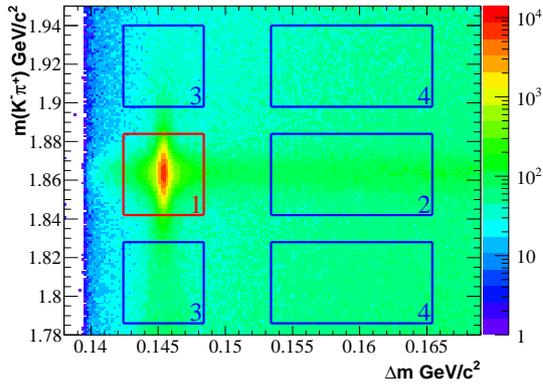}
\caption{m(\Km\pip) vs. $\Delta m$ scatter plot of the data sample. Signal region is identified
by the red box and the blue lines show the sidebands. The numbers identify the events category.}
\label{fig:mDvsDm}
\end{center}
\end{figure}

As shown in Fig.~\ref{fig:mDvsDm}, four categories of events can be recognized after the reconstruction:
\begin{enumerate}
        \item signal: real \Dz and $\pip_s$ from \Dstarp decay.
        \item Missed $\pip_s$: real \Dz\to\Km\pip decay that may or may not have
        come from a \Dstarp, combined to a \pip from combinatoric.
        \item Missed \Dz: a mis-reconstructed \Dz with a real $\pip_s$ from \Dstarp.
        This is mostly \Dz\to\Km\Kp, \Dz\to\Km\pip\piz, \Dz\to\pip\pim or cases where
        the kaon and pion assignments have been swapped.
        \item Combinatoric background: neither \Dz or $\pip_s$ are correctly
        reconstructed from a \Dstarp decay.
\end{enumerate}

%All the three backgrounds incoherently contribute to the number of events in the signal region.
%In order to estimate this contribution the control sample is used. This sample is dominated 
%by background events, since no requirements are made on kaon particle identification.           

The amount of combinatoric and real \Dz, fake $\pip_s$ background in the signal
region is estimated using the re-normalized distribution of the control sample
\Dz sidebands in the $\Delta m=m(\Km\pip\pip_s)- m(\Km\pip)$ signal region.
The scale factor needed for going from $\Delta m$ sideband to the $\Delta m$ signal
region is taken from the control sample itself.
This background subtraction procedure has been carried out for the
$\cos\theta^*$ distribution for each bin of $p^*(\Dstar)$ using the following steps:

\begin{enumerate}
        \item consider the {\it no PID} sample in the m($K^{-}\pi^{+}$)
          sideband regions; divide the number of events in the $\Delta                                                                                                      
          m$ signal region by the number of events in the $\Delta m$
          sideband to get the scale factor to go from sideband to signal
          in $\Delta m$;
        \item in the {\it good PID} sample scale the m($K^{-}\pi^{+}$)
          spectrum in the $\Delta m$ sideband region using the factor
          obtained in the previous step; then integrate the resulting
          m($K^{-}\pi^{+}$) spectrum to get the factor to rescale
          background;
        \item use the factor measured in step 2 to rescale the
          interesting distribution ($\cos\theta^*$) obtained from the
          events of the {\it no PID} sample in $\Delta m$ signal,
          m($K^{-}\pi^{+}$) sideband region (category 4);
        \item subtract the distribution obtained in step 3 from the same
          distribution retrieved from {\it good PID} sample in signal
          region.
\end{enumerate}

This procedure has been carried out for the $\cos\theta^*$ distribution
for each bin of $p^*(\Dstar)$. All the histograms have been then fit to
the convolution of Eqs.~\ref{eq:SoftPiCosTh} and~\ref{eq:SoftPiEff} to determine
the parameters of the efficiency function $p_0$ and $\delta$. Measuring
$p_0$ and $\delta$ for data and MC, we can evaluate the systematic error
using Eq.~\ref{eq:SoftPiSyst}. The fit makes use of a global $\chi^2$ defined
as

\begin{equation}
\chi^2 = \sum_{l,k}\frac{\left( D_{lk} - S_{lk} \right)^2}{\sigma^2_{D_{lk}}},
\label{eq:SoftPiChi2}
\end{equation}
where $k$ is the index referring to one of the 8 $p^*(\Dstar)$ regions, $l$ refers
to one of the 16 bins of the $cos\theta^*$ histogram in that region.
$D_{lk}$, $\sigma_{D_{lk}}$ and $S_{lk}$ are the number of events observed in
the bin, its error and the number of events expect by the fit model, respectively.
The expression of the fit model is
\begin{equation}
S_{lk} = \sum_{i,j}\epsilon(p_{ij}; p_0, \delta)N_{k}(1+\alpha_k\cos\theta^*_{i})
\label{eq:Slk}
\end{equation}
where $i$ indicates the bin of the $\cos\theta^*$ distribution in the $k^{th}$
$p^*(\Dstar)$ region, and $j$ is one of the 10 bins of the detailed
distribution within the range of momentum considered in the $k^{th}$
$p^*(\Dstar)$ region.

\begin{figure}[ht]
\centering
\includegraphics[width=0.45\textwidth]{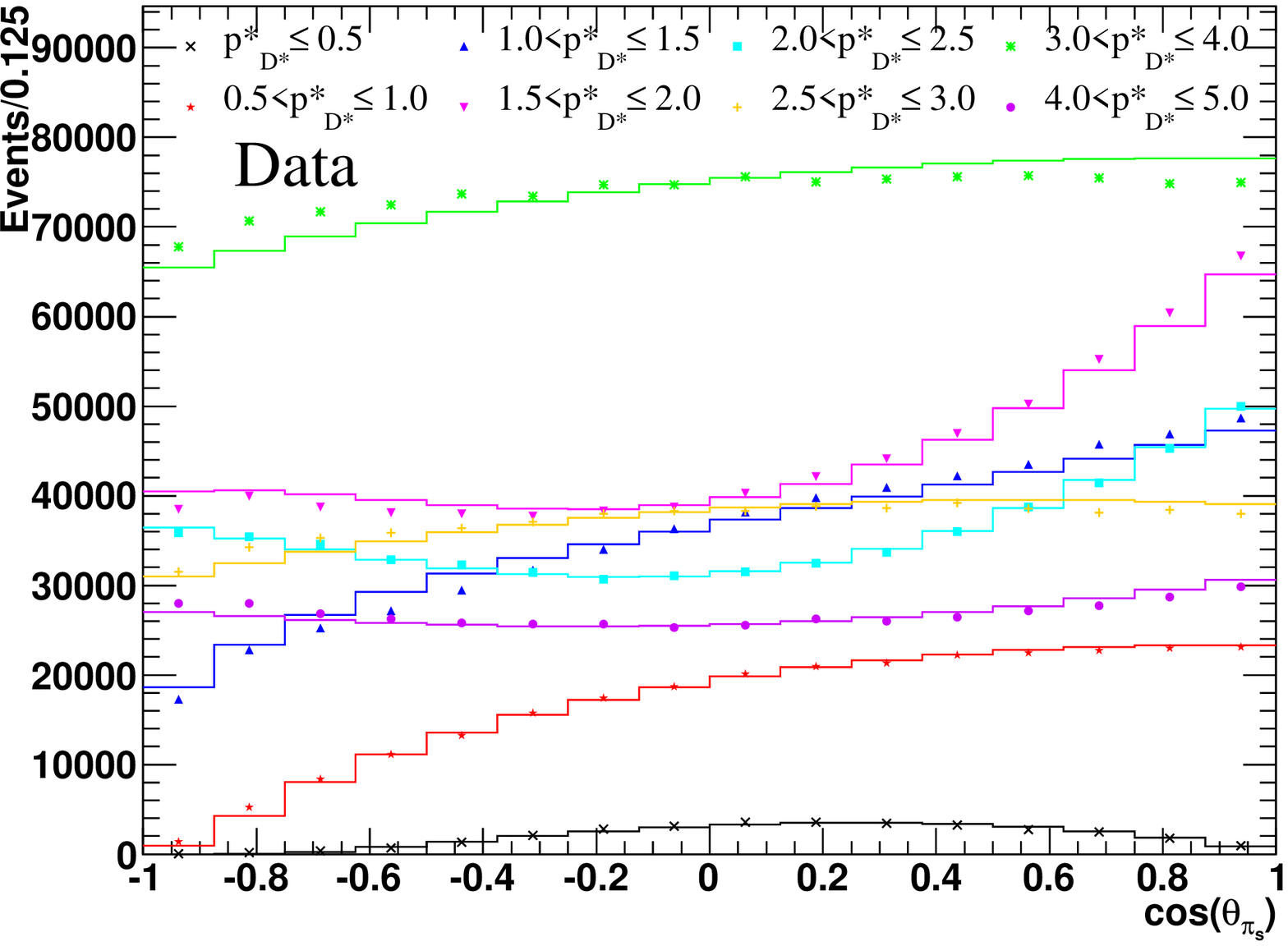}
\includegraphics[width=0.45\textwidth]{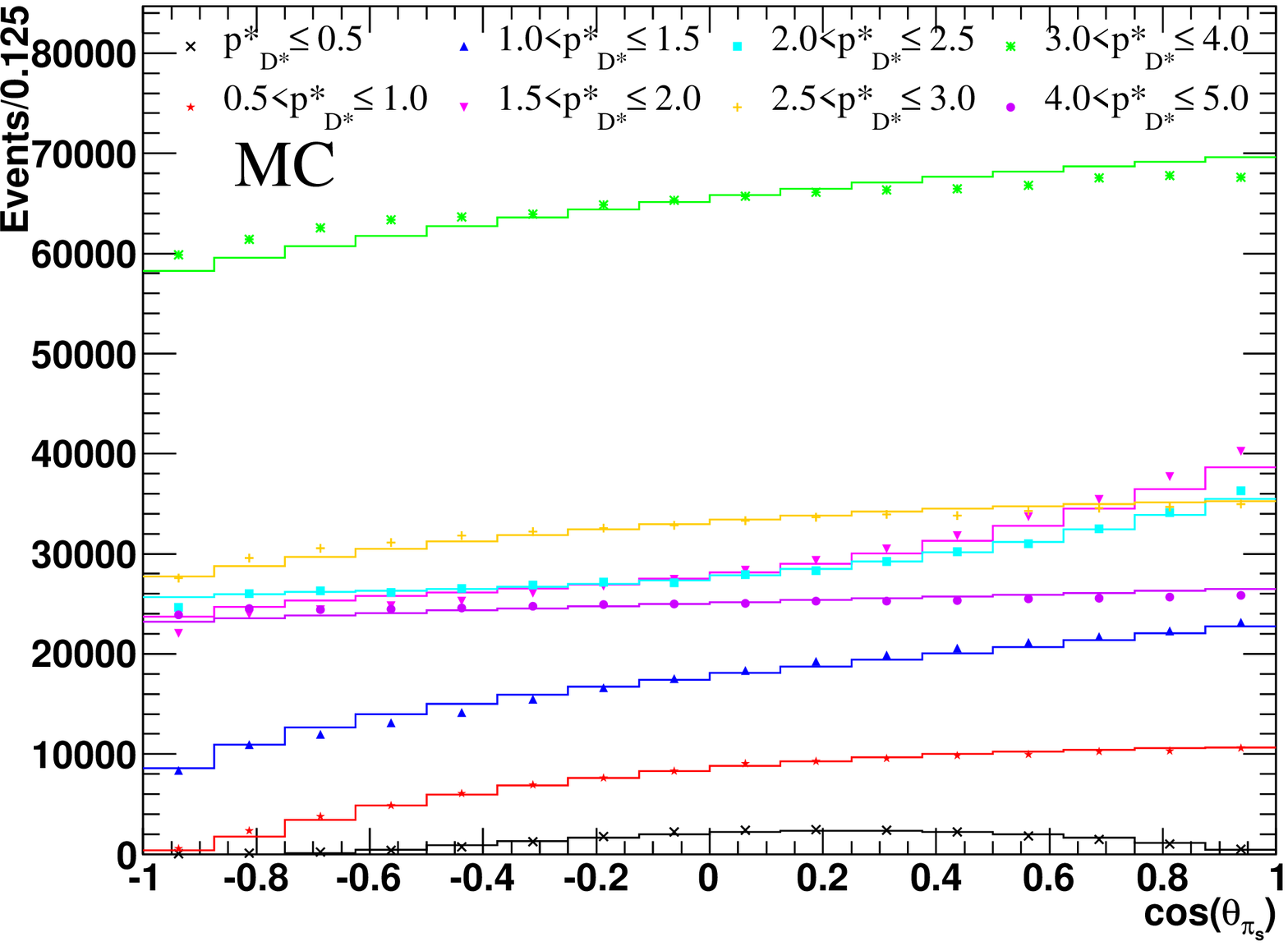}
\caption{Fit of the model distribution to data (top) and MC (bottom).
  In both the plots, the measured data/MC event ratios are represented
  by the dots, while the fit results are shown using a line histogram.
  The distributions of $\cos\theta^*$ in the different ranges of
  $p^*(\Dstar)$ are shown using different colors, as outlined in the
  legend (the $p^*(\Dstar)$ values are measured in \gevc).  The fit to
  data returns $\chi^2/n_{dof} = 1.34$; the one to MC gives
  $\chi^2/n_{dof} = 0.71$.}
\label{fig:SoftPiFit}
\end{figure}

The number of floating parameters in the fit are 18: 8 normalization
factors $N_i$, 8 $\alpha_i$ (one for each bin) from Eq.~\ref{eq:SoftPiCosTh},
and $\delta$ and $p_0$ from Eq.~\ref{eq:SoftPiEff}. The fit has been made both
to data and MC, giving the results shown in Fig.~\ref{fig:SoftPiFit} and
Tab.~\ref{tab:SoftPiFit}.

\begin{table}[htb]\footnotesize
\begin{center}
\caption{Fit results on data and MC.}
\begin{tabular}{cccc}
\toprule
Parameter       & MC                                  & Data \\
\midrule
 $\delta$        &$13.77 \pm 0.18$              &$14.54\pm0.15$\\
 $p_0$          &$27.82\pm0.21$\mevc            &$27.01\pm0.14$\mevc\\
 $N_1$          &$1284645\pm1784$               &$185581\pm1927$\\
 $\alpha_1$     &$-9.88\pm0.12\times10^{-1}$
&$-9.30\pm0.10\times10^{-1}$\\
 $N_2$          &$452849\pm3524$
&$970429\pm5694$\\
 $\alpha_2$     &$-8.25\pm1.41\times10^{-2}$
&$-9.75\pm0.86\times10^{-2}$\\
 $N_3$          &$738162\pm4631$
&$1482799\pm7329$\\
 $\alpha_3$     &$7.12\pm0.89\times10^{-2}$
&$8.95\pm0.60\times10^{-2}$\\
 $N_4$          &$1023973\pm5400$               &$1527424\pm6434$\\
 $\alpha_4$     &$2.53\pm0.07\times10^{-1}$
&$5.19\pm0.06\times10^{-2}$\\
 $N_5$          &$894908\pm4092$
&$1093406\pm4040$\\
 $\alpha_5$     &$1.90\pm0.06\times10^{-1}$
&$5.13\pm0.07\times10^{-1}$\\
 $N_6$          &$937073\pm3729$
&$1050245\pm3390$\\
 $\alpha_6$     &$-1.93\pm0.49\times10^{-2}$
&$-6.36\pm0.44\times10^{-2}$\\
 $N_7$          &$1738326\pm5687$               &$1948264\pm5116$\\
 $\alpha_7$     &$0.98\pm3.43\times10^{-3}$
&$-2.78\pm0.32\times10^{-2}$\\
 $N_8$          &$626949\pm1881$
&$665218\pm1678$\\
 $\alpha_8$     &$1.46\pm0.55\times10^{-2}$
&$1.73\pm0.06\times10^{-1}$\\
\bottomrule
\end{tabular}
\label{tab:SoftPiFit}
\end{center}
\end{table}

\begin{figure}[ht]
\centering
\includegraphics[width=0.45\textwidth]{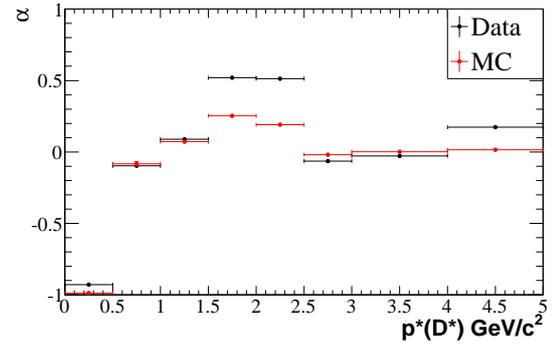}
\caption{Comparison of the fit results on $\alpha$ in the 8 bins of $p^*(\Dstar)$ for data (black) and Monte Carlo (red).
The difference observed in the $4^{th}$, $5^{th}$ and $8^{th}$ bins is due to the slightly different
helicity distribution for data and MC in these $p^*(\Dstar)$ ranges.}
\label{fig:SoftPiFitAlpha}
\end{figure}

\begin{figure}[htb]
\begin{center}
\includegraphics[width=0.45\textwidth]{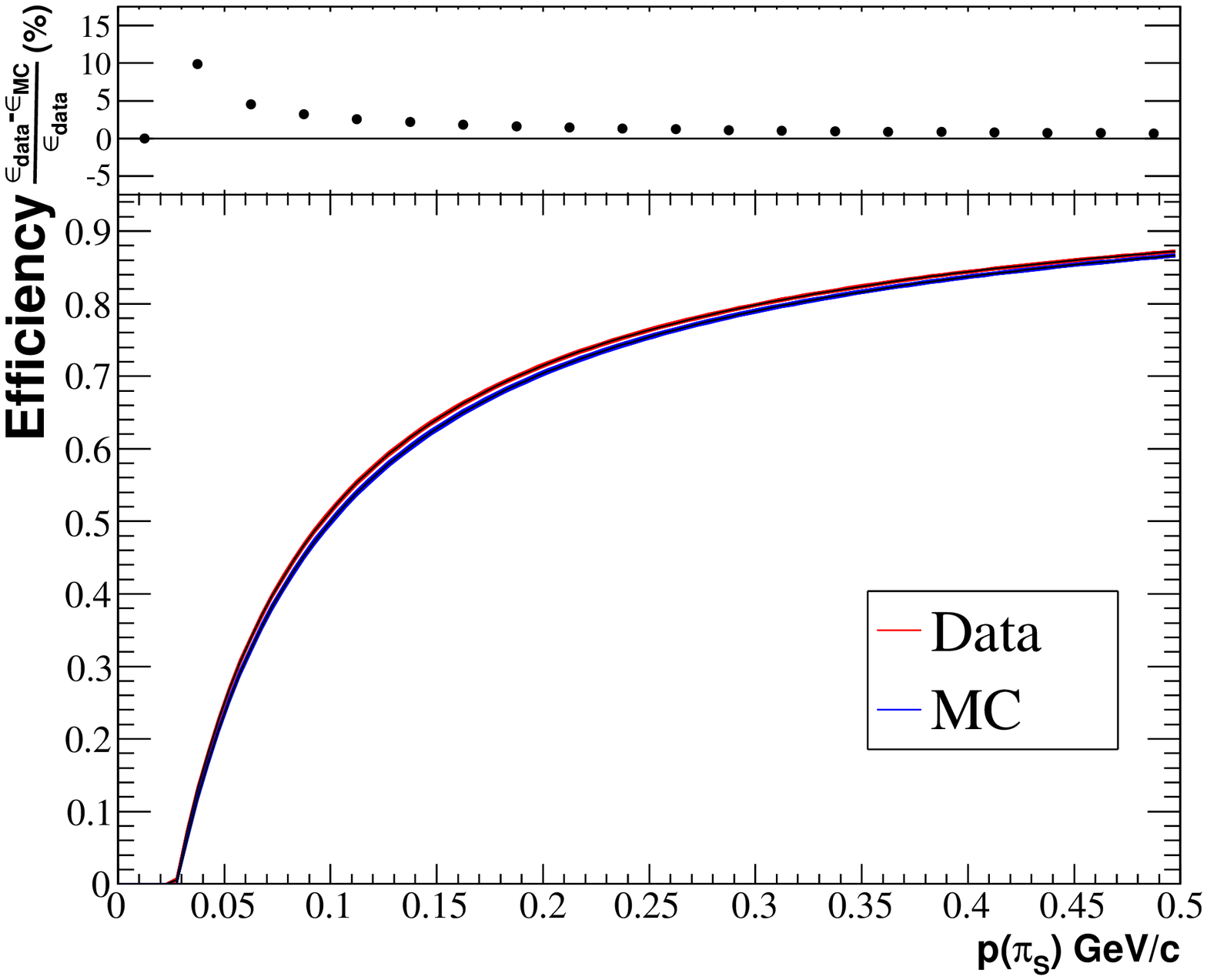}
\caption{Soft pion reconstruction efficiency functions obtained from the
  fit to data (red) and MC (blue).  Both the efficiency functions are
  shown together with the functions obtained by varying the central
  values of the fit parameters $p_0$ and $\delta$ by 1$\sigma$.  The
  curve obtained using the central values is drawn in black. On top of
  the curves, the distribution of the relative difference between data
  and MC is shown.}
\label{fig:SoftPiEff}
\end{center}
\end{figure}

Finally, the efficiency functions are compared in
Fig.~\ref{fig:SoftPiEff}. Please note that these distributions include
acceptance. The method shown herein does not allow to disentangle the
acceptance from the soft pion efficiency. The systematic uncertainty estimated from
Eq.~\ref{eq:SoftPiSyst} for the low $p_T$ tracks is $\sigma_{syst} = 1.54\%$.

\section{$K_S^0$ reconstruction efficiency measurement}

A significant number of analyses in \babar\ involve the reconstruction of
the decay $K_S^0 \to \pi^+ \pi^-$, where the two charged pions belong to
the list CT of all reconstructed tracks in the event. The
track reconstruction efficiency for charged tracks originating within
15 mm in XY from the beam spot is studied by the other methods presented earlier
in the paper. However, most of the $K_S^0$'s decay outside this 15 mm
radius, making it necessary to understand the $K_S^0$ daughter reconstruction
efficiency in data and MC. 

The reconstruction efficiency of the $K_S^0$ daughters depends on the
$K_S^0$ transverse momentum, $p_T$, polar angle, $\theta_{LAB}$ and
transverse ($XY$) flight distance, $d_{XY}$, which is computed as the
distance between the primary vertex of the event and the refitted
$K_S^0$ decay vertex.

The general strategy is to subdivide the data and MC events into a large
number of samples by choosing an appropriate binning in these variables,
determine the number of $K_S^0$'s in each bin in data and MC and, for
each of the momentum and polar angle ranges, normalize the ratio of its
value in the first bin in $d_{XY}$, where all tracking effects are
understood to 1.000 by definition, with
no associated error other than the systematic uncertainty per track, as
discussed in Section~\ref{sec:tau31eff}. Bin sizes are optimized to
ensure a sufficient number of events in each bin, with 4 bins in $p_T$
(0.0 - 0.4 - 1.0 - 2.0 - 4.0 GeV/c), 8 bins in $\theta_{LAB}$
($7.0^\circ - 25.6^\circ - 44.25^\circ - 62.88^\circ - 81.5^\circ -
100.13^\circ - 118.75^\circ -137.38^\circ - 156.0^\circ$) and 10 bins in
$d_{XY}$ (0.0 - 0.3 - 1.3 - 2.78 - 3.2 - 4.0 - 5.4 - 9.1 - 11.4 - 23.6 -
40.0 cm).  The binning in $d_{XY}$ roughly reflects the structure of the
\babar\ detector and the bins are numbered from 0 to 9, bin 1 being the
normalization bin.  The normalized ratio of the data and MC as a
function of different bins is provided as a correction factor for the
$K_S^0$ daughter reconstruction efficiency.  In order to reduce the uncertainty
from the imperfect simulation of the random track background or the
potential differences between the $K_S^0$ quality cut efficiencies in
data and MC, we remove the immediate vicinity of the event's primary
vertex which is 3 mm in XY from the first (normalization) bin.

\begin{figure}[htb]
\begin{center}
\includegraphics[width=0.4\textwidth]{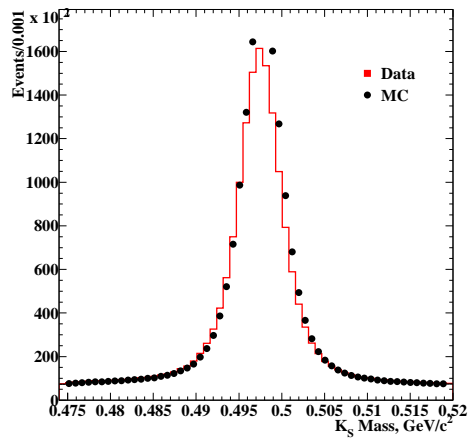}
\includegraphics[width=0.4\textwidth]{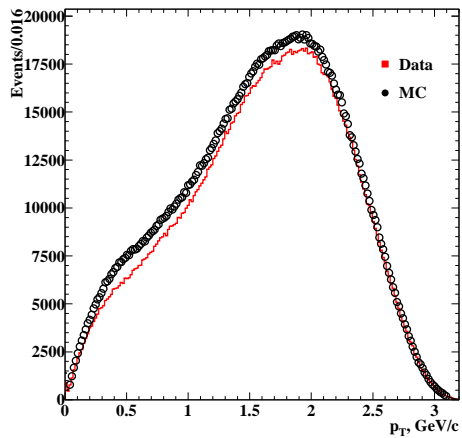}
\caption{$K_S^0$ mass (top) and transverse momentum (bottom) for data
  and MC. MC is normalized to the data luminosity.}
\label{fig:ks}
\end{center}
\end{figure}

\begin{figure}[htb]
\begin{center}
\includegraphics[width=0.4\textwidth]{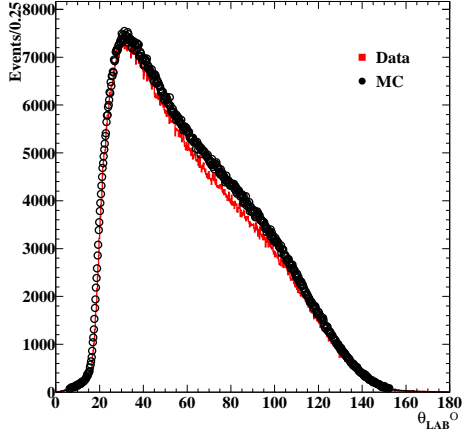}
\includegraphics[width=0.4\textwidth]{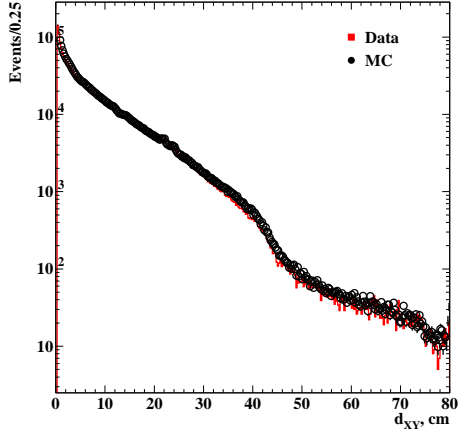}
\caption{$K_S^0$ polar angle (top) and transverse flight length (bottom)
  for data and MC. MC is normalized to the data luminosity.}
\label{fig:ks1}
\end{center}
\end{figure}

Events of interest are selected by looking for the $B \to h^+ h^- K_S^0$
(with $h=\pi, K$) decays in the data and MC samples. The MC sample includes events
from generic B decays, light quark events (u,d,s,c) and $\tau^{+} \tau^{-}$ decays.
The $K_S^0$ is reconstructed from two oppositely charged tracks, the invariant mass of
which is required to be within 25 MeV/$c^2$ of the PDG value of the $K_S^0$ mass
($m_{k^0_s}$ = 497.614 $\pm$ 0.024 MeV/$c^2$)~\cite{pdg}.  The two oppositely
charged tracks must originate from a common vertex and the fit is
required not to fail. The event is required to have at least five GT
tracks, two of which are oppositely charged GT tracks that when combined
with the $K_S^0$ candidate to form an object with $m_{ES} > 5.19$
GeV/$c^2$ and $|\Delta E| < 0.3$ GeV. About 93\% of these events come
from the light quark (udsc) continuum; the contribution from $\tau^{+}
\tau^{-}$ production is about 2.5\% and only about 3.5\% of candidates
arise from B decays, most of which are random track combinations. 
Figures~\ref{fig:ks} and ~\ref{fig:ks1} show the data and MC comparison of the
$K_S^0$ mass, $p_T$, $\theta_{LAB}$ and $d_{XY}$ distributions for the reconstructed
$K_S^0$ candidates. 

%Since the two CT tracks that the $K_S^0$ candidates are composed of are
%unlikely to also be on the GT list, there are usually 4 or 5, and never
%less than 3 non-$K_S^0$ candidates in the event, which ensures a high
%quality of primary vertex reconstruction. 

%To determine the numbers of $K_S^0$'s in each of the bins in data and
%MC, and for each of the momentum and polar angle ranges, we normalize
%the ratio of these numbers to its value in the first bin in $d_{XY}$,
%where all tracking effects are well understood and thus the ratio is
%1.000 by definition, with no associated error other than the 0.5\%
%systematic uncertainty per track.

%We eliminate most of the combinatorial (random-track) background to
%$K_S^0 \to \pi^+ \pi^-$ by removing tracks in the immediate vicinity of
%the event's primary vertex, i.e., within 3 mm in XY, from the first
%(normalization) bin. 

To determine the number of $K_S^0$'s in each of the bins in 
data and MC, the $K_S^0$ mass distributions in each of the bins are fitted with a sum
of a double Gaussian and a constant background. The constant background, determined from the sideband
regions in the $K_S^0$ mass distribution, [0.476, 0.485] U [0.511, 0.520] GeV/$c^2$, for each bin,
is then subtracted to determine the number of $K_S^0$ in each bin. The limited statistics 
in a large fraction of bins makes it impractical to allow the slope of the background to float. The binned 
maximum log likelihood method is used since the default $\chi^2$-minimization method is less appropriate in this
study as it systematically, in a statistics-dependent way, underestimates the number of events in each bin.

We define the values of the normalized ratios $R_{ijk}$ and
the uncertainties $\sigma_{R_{ijk}}$, where the indices $i$, $j$ and $k$
stand for the number of the $p_T$, $\theta_{LAB}$ and $d_{XY}$ bins
respectively, to be

\begin{equation}
R_{ijk} = (N_{ijk}/M_{ijk}) / (N_{ij1}/M_{ij1}) 
\label{eq1}
\end{equation}

\begin{equation}
\sigma_{R_{ijk}}=R_{ijk}\sqrt{\left(\sigma_{N_{ijk}}/N_{ijk}\right)^2+\left(\sigma_{M_{ijk}}/M_{ijk}\right)^2}
\label{eq2}
\end{equation}

where $N_{ijk}$ and $\sigma_{N_{ijk}}$ are the numbers of $K_S^0$'s and their uncertainties in data and $M_{ijk}$ and $\sigma_{M_{ijk}}$
are the numbers of $K_S^0$'s and their uncertainties in MC. We also take into account the difference in the $K_S^0$ mass resolutions in data and MC 
by performing numerical integration to determine the efficiencies of the $|m_{\pi^+ \pi^-} - m_{K_S^0}| < 12$ MeV/$c^2$ cut in data and MC. 
The efficiency corrected normalized ratio, $R_{ijk}$, also called the correction factor, is computed for each bin and also as a function of several
different sets of $K_S^0$ quality cuts (cuts on the $K_S^0$ mass, 3-D or $XY$ flight length or its significance, the 3-D or $XY$ angle, $\alpha$ between the 
$K_S^0$ momentum and the line connecting the $K_S^0$ decay vertex and the primary vertex of the event, and, in one instance, a cut on the 
$K_S^0 \to \pi^+ \pi^-$ vertexing quality), which also enables us to study the systematic uncertainties associated with a correction factor.

%To take into account the difference in the $K_S^0$ mass resolutions in data and MC, we perform numerical integration to determine the efficiencies
%of the $|m_{\pi^+ \pi^-} - m_{K_S^0}| < 12$ MeV/$c^2$ cut in data ($\epsilon_{ijk}$) and MC ($\xi_{ijk}$). Then the 
%$K_S^0$ efficiency correction in bin $ijk$ and their respective uncertainties $\sigma_{R_{ijk}^{eff}}$ are given by

%\begin{equation}
%R_{ijk}^{eff} = R_{ijk} (\epsilon_{ijk}/\xi_{ijk}) 
%\label{eq3}
%\end{equation}

%\begin{equation}
%\sigma_{R_{ijk}^{eff}}= \sigma_{R_{ijk}} (\epsilon_{ijk}/\xi_{ijk}) 
%\label{eq4}
%\end{equation}

To get the overall correction in the $K_S^0$ daughter reconstruction efficiency in an analysis, we apply the correction factors to 
the signal MC. If $H_{ijk}$ is the number of events in the signal MC sample that falls within 
the bin (ijk), the relative weight of a $K_S^0$ reconstruction efficiency correction table element 
$R_{ijk}$ is $H_{ijk}$ / $H_{tot}$, where $H_{tot}$ = $\Sigma H_{ijk}$ and its statistical uncertainty 
is $\sqrt{(H_{ijk}) / H_{tot}}$. The central value of the overall data/MC 
efficiency ratio is simply given by

\begin{equation}
R=\frac{1}{H_{tot}}\sum_{ijk}H_{ijk}C_{ijk} = \sum_{ijk}R_{ijk}C_{ijk}
\label{eqR}
\end{equation}

Calculation of the statistical uncertainty on this number is slightly
non-trivial since we have to take into account the fact that the
statistical uncertainty on the normalization bin ratio,
$\sigma_{R_{ij1}}$, influences the entire row (ij). Substituting
$R_{ijk}$ in Eq.~\ref{eqR} with the expression in Eq.~\ref{eq1} and
differentiating the resulting expression with respect to each of the
variables that enter it while using $R_{ij1}$ = 1, we obtain

\begin{eqnarray}
\sigma_{R}=\frac{1}{H_{tot}}\sum_{ij}\bigg\{ \sum_{k} H_{ijk} R_{ijk}^2
  + \nonumber\\ \sum_{k \not = 1} \{H_{ijk} \sigma_{R_{ijk}}\}^2 +
  \{\sum_{k \not = 1} H_{ijk} R_{ijk} \sigma_{R_{ij1}}\}^2
  \bigg\}^{\frac{1}{2}}
\label{eqsigR}
\end{eqnarray}

where the first term reflects the finite size of the signal MC sample
used in the study and is generally the smallest, the second term
reflects the statistical uncertainties on the number of $K_S^0$'s in
bins other than the normalization bin, and the third term, the dominant
one, reflects the dependence on the statistical precision of $R_{ij1}$
of the correction factors in each of the bins $R_{ijk}$ , $k \not = 1$.

The $K_S^0$ correction factors are applied to signal MC for the decay
modes $B^0 \to \phi K_S^0$ and $B^0 \to \pi^+ D^- (D^- \to K_S^0
\pi^-)$, which provide $K_S^0$ spectra representative of most cases of
$K_S^0$ used in \babar~ analyses, to determine the overall correction
factor and its statistical error. The above exercise is repeated for
several sets of $K_S^0$ quality cuts, from none to tight, and for three
different binning approaches. Half of the largest deviation 
in the $K_S^0$  correction factors for different $K_S^0$ quality cuts is considered to 
be the systematic uncertainty associated with $K_S^0$ daughter reconstruction efficiency.
For these modes we are able to determine the ratio of the data
and MC $K_S^0$ daughter reconstruction efficiency to be about 99.5\% and with a
statistical error of $\sim 0.4\%$ and a systematic uncertainty of $\sim$0.7\%.

\section{Conclusion}
In conclusion, we studied the track reconstruction efficiency of charged
particles in \babar ~over a wide range of momentum, polar angle, and
track separation.  Our results come from
several different control samples, which are observed to be self-consistent,
and well modeled
in MC.
The overall reconstruction efficiency for isolated tracks is found to be
consistent with MC predictions.
We also measured the charge asymmetry in the track reconstruction, which was found be
consistent with zero. Any observed difference between data and MC in the
track reconstruction efficiency could be considered as a source of
systematic uncertainty in all the physics analyses in \babar.  For
physics analyses with low multiplicity and similar topology to the Tau31
decays and for charged tracks with momentum greater than 180 MeV/c, the
results from the Tau31 study should be used; for other B and D decays,
the results from the ISR study could be considered. For charged tracks
with momenta less than 180 MeV/c, an additional systematic uncertainty
of 1.54\% per track should be applied. The $K_S^0$ reconstruction
efficiency in data and MC is found to be a function of $K_S^0$ momentum,
polar angle and transverse flight distance, which needs to be considered
for $K_S^0$ reconstruction in \babar.

\section{Acknowledgements} 
We are grateful for the extraordinary contributions of our
\pep2\ colleagues in achieving the excellent luminosity and machine
conditions that have made this work possible.  The success of this
project also relies critically on the expertise and dedication of the
computing organizations that support \babar.  The collaborating
institutions wish to thank SLAC for its support and the kind hospitality
extended to them.  This work is supported by the US Department of Energy
and National Science Foundation, the Natural Sciences and Engineering
Research Council (Canada), the Commissariat \`a l'Energie Atomique and
Institut National de Physique Nucl\'eaire et de Physique des Particules
(France), the Bundesministerium f\"ur Bildung und Forschung and Deutsche
Forschungsgemeinschaft (Germany), the Istituto Nazionale di Fisica
Nucleare (Italy), the Foundation for Fundamental Research on Matter (The
Netherlands), the Research Council of Norway, the Ministry of Education
and Science of the Russian Federation, Ministerio de Ciencia e
Innovaci\'on (Spain), and the Science and Technology Facilities Council
(United Kingdom).  Individuals have received support from the
Marie-Curie IEF program (European Union), the A. P. Sloan Foundation
(USA) and the Binational Science Foundation (USA-Israel).

\bibliographystyle{model1-num-names}

\begin{thebibliography}{00}
\bibitem{conj}
Charge-conjugate reactions are implied throughout this paper except for
section~\ref{TrackingChargeAsymmetry}.
\end{thebibliography}
%\begin{thebibliography}{99}

\end{document}